\documentclass[aip,jcp,twocolumn,10pt,floatfix]{revtex4-1}

\usepackage[OT1]{fontenc}
\usepackage[utf8]{inputenc}
\usepackage{graphicx}
\usepackage{braket}
\usepackage{siunitx}
\usepackage{amsmath}
\usepackage{amssymb}
\usepackage{bm}
\usepackage{booktabs}
\usepackage[capitalize]{cleveref}

\usepackage{color}

\begin{document}

\title{Vibrationally resolved electronic spectra including vibrational pre-excitation:
       Theory and application to VIPER spectroscopy}

\author{Jan von Cosel}
\affiliation{Institute of Physical and Theoretical Chemistry, Goethe University Frankfurt,
             Max-von-Laue-Str. 7, 60438 Frankfurt, Germany}

\author{Javier Cerezo}
\affiliation{Departamento de Qu\'imica F\'isica, Universidad de Murcia, E-30071 Murcia, Spain}

\author{Daniela Kern-Michler}
\affiliation{Institute of Biophysics, Goethe University Frankfurt, Max-von-Laue-Str. 1, 60438
             Frankfurt, Germany}

\author{Carsten Neumann}
\affiliation{Institute of Biophysics, Goethe University Frankfurt, Max-von-Laue-Str. 1, 60438
             Frankfurt, Germany}

\author{Luuk J. G. W. van Wilderen}
\affiliation{Institute of Biophysics, Goethe University Frankfurt, Max-von-Laue-Str. 1, 60438
             Frankfurt, Germany}

\author{Jens Bredenbeck}
\affiliation{Institute of Biophysics, Goethe University Frankfurt, Max-von-Laue-Str. 1, 60438
             Frankfurt, Germany}

\author{Fabrizio Santoro}
\email[Author to whom correspondence should be addressed: ]{fabrizio.santoro@pi.iccom.cnr.it}
\affiliation{Consiglio Nazionale delle Ricerche -- CNR, Istituto di Chimica dei Composti Organo
             Metallici (ICCOM-CNR), UOS di Pisa, Via G. Moruzzi 1, I-56124 Pisa, Italy}

\author{Irene Burghardt}
\email[Author to whom correspondence should be addressed: ]{burghardt@chemie.uni-frankfurt.de}
\affiliation{Institute of Physical and Theoretical Chemistry, Goethe University Frankfurt,
             Max-von-Laue-Str. 7, 60438 Frankfurt, Germany}

\date{\today}

\begin{abstract}
  Vibrationally resolved electronic absorption spectra including the effect of
  vibrational pre-excita\-tion are computed in order to interpret and predict
  vibronic transitions that are probed in the Vibrationally Promoted
  Electronic Resonance (VIPER) experiment [L. J. G. W. van Wilderen et al.,
    {Angew.\ Chem.\ Int.\ Ed.} {\bf 53}, 2667 (2014)]. To this end, we employ
  time-independent and time-dependent methods based on the evaluation of
  Franck-Condon overlap integrals and Fourier transformation of time-domain
  wavepacket autocorrelation functions, respectively. The time-independent
  approach uses a generalized version of the \emph{FCclasses} method
  [F.\ Santoro et al., {J.\ Chem.\ Phys.} {\bf 126}, 084509 (2007)]. In the
  time-dependent approach, autocorrelation functions are obtained by
  wavepacket propagation and by evaluation of analytic expressions, within the
  harmonic approximation including Duschinsky rotation effects. For several
  medium-sized polyatomic systems, it is shown that selective pre-excitation
  of particular vibrational modes leads to a redshift of the low-frequency
  edge of the electronic absorption spectrum, which is a prerequisite for the
  VIPER experiment. This effect is typically most pronounced upon excitation
  of {\color{black} modes that are significantly displaced during the electronic
      transition, like} ring distortion modes within an aromatic $\pi$-system. Theoretical
  predictions as to which modes show the strongest VIPER effect are found to
  be in excellent agreement with experiment.
\end{abstract}

\maketitle

\section{Introduction}
\label{sec.intro}

\begin{figure}[b]
    \begin{minipage}{0.47\columnwidth}
        \includegraphics[width=0.8\textwidth]{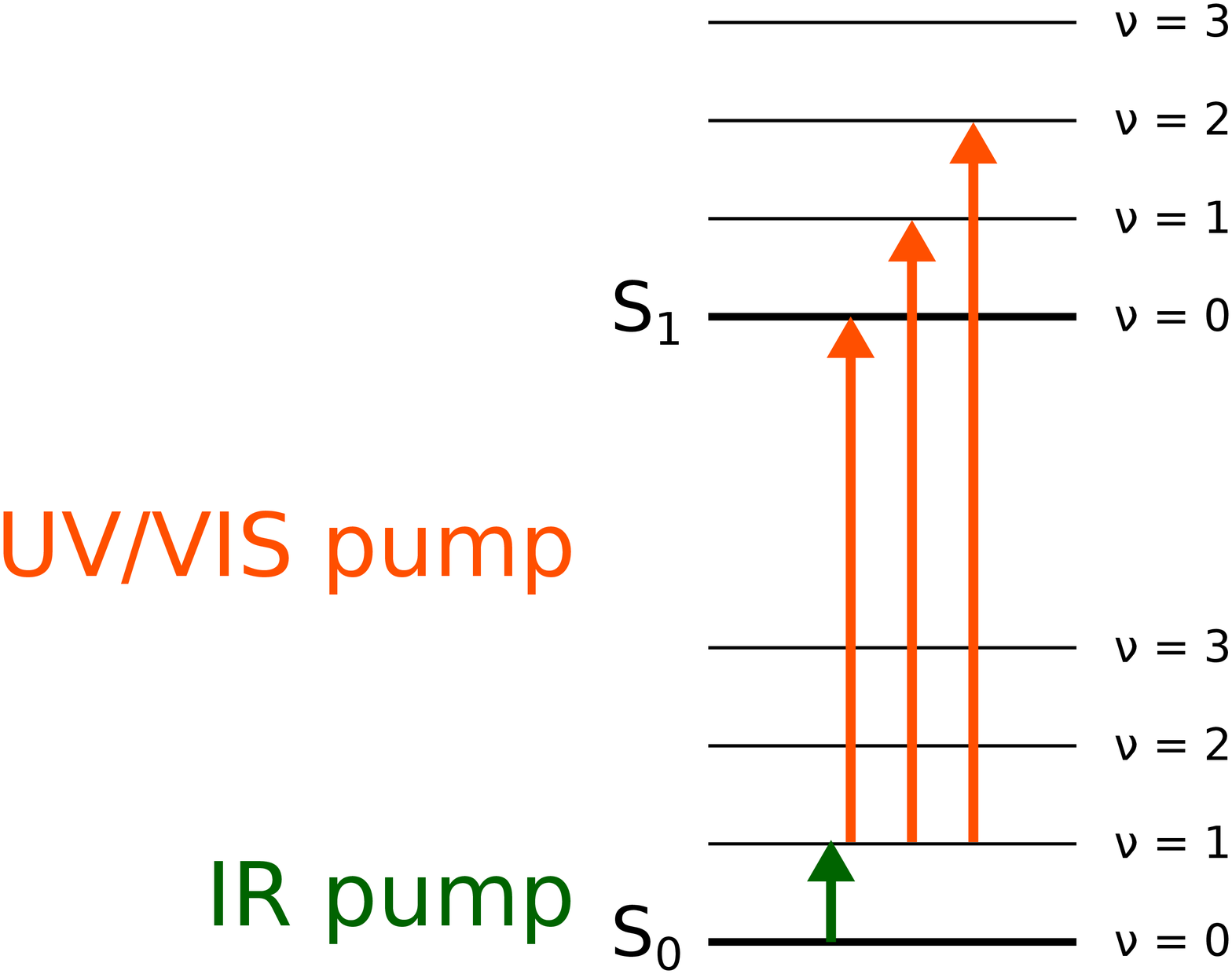}
    \end{minipage}
    \begin{minipage}{0.47\columnwidth}
        \includegraphics[width=0.9\textwidth]{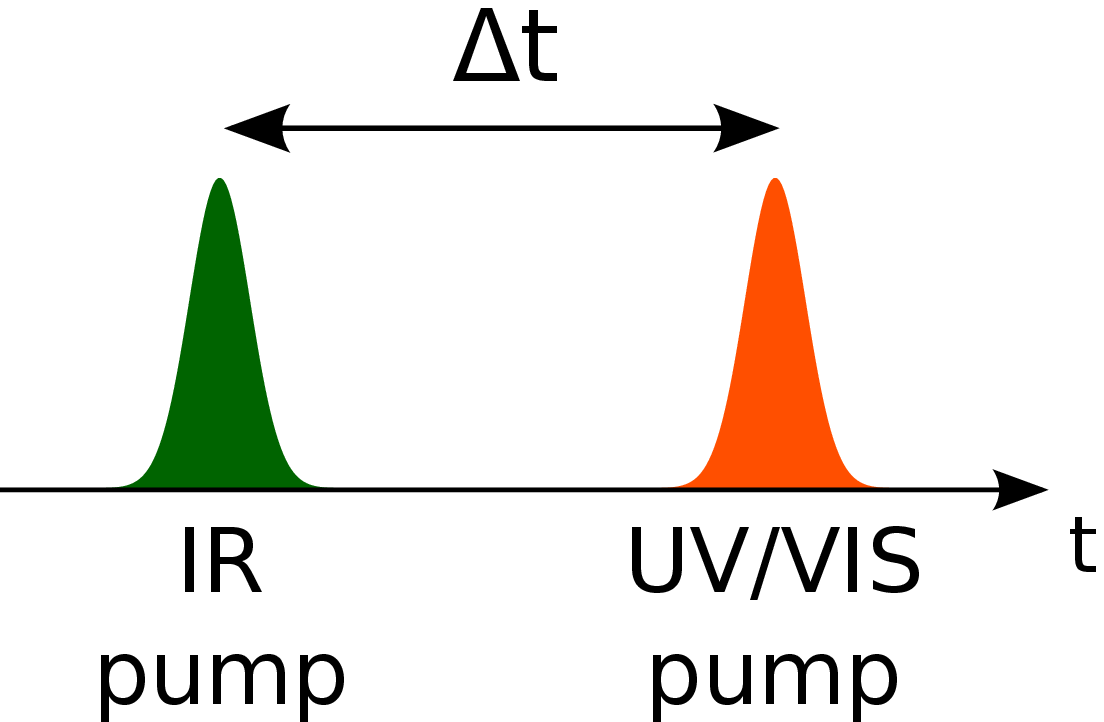}
    \end{minipage}
    \caption{Left: The mixed IR/VIS VIPER excitation scheme.
             Right: Time-domain VIPER excitation sequence.
            }\label{fig:vipseq}
\end{figure}

Combined electronic-vibrational spectroscopies pave the way for new strategies
of probing and controlling molecular systems.\cite{vanwilderen_ultrafast_2015}
This is exemplified by the recently introduced Vibrationally Promoted
Electronic Resonance (VIPER) experiment
\cite{vanwilderen_ultrafast_2015,bredenbeck_viper2014_int} where selective
infrared (IR) excitation in the electronic ground state precedes visible (VIS)
excitation, as sketched in Fig.\ \ref{fig:vipseq}. The VIPER pulse sequence
was originally designed in the context of chemical
exchange\cite{bredenbeck_viper2014_int}, but also lends itself to inducing
selective cleavage of photolabile protecting groups through the use of
selective IR excitation and isotope substitution. For example, substitution of
{$^{12}$C} by {$^{13}$C} leaves the electronic absorption spectrum
{\color{black} of bright transitions} virtually
unchanged while the frequencies of vibrational modes can be altered
significantly. Different isotopomers can therefore be distinguished by their
IR spectra and selectively excited using narrow-band IR pulses
\cite{kern_steering_inprep}.

In the present work, we compute vibrationally resolved electronic absorption
spectra including the effect of vibrational pre-excitation, in order to
predict VIPER-active modes for several polyatomic chromophores, some of which
have already been experimentally investigated. Notably, the laser dye
Coumarin 6 is studied, which has served for demonstration of VIPER's
capability to measure exchange beyond the vibrational $T_1$
lifetime\cite{vanwilderen_ultrafast_2015}. Furthermore, [7-(diethylamino)coumarin-4-yl]methyl-azide
(DEACM-N$_3$) and \emph{para}-Hydroxyphenacyl thiocyanate (\emph{p}HP-SCN) are investigated, which
are model systems for photo-cleavable caging
groups\cite{klan_photoremovable_2012,Givens97,Bredenbeck2017}, with the azide and thiocyanate groups
representing the relevant leaving groups. {\color{black} We aim} to establish a
general understanding of which properties favor large spectral shifts that
permit effective VIPER excitation. While we do not simulate the complete
{\color{black} VIPER pump-probe sequence\cite{vanwilderen_ultrafast_2015,bredenbeck_viper2014_int}}, our analysis of the
vibronic spectra resulting from the combined IR-pump/VIS-pump steps, is
suitable for the {\color{black} prediction} of the relevant VIPER shifts.

\begin{figure}
    \includegraphics[width=0.7\columnwidth]{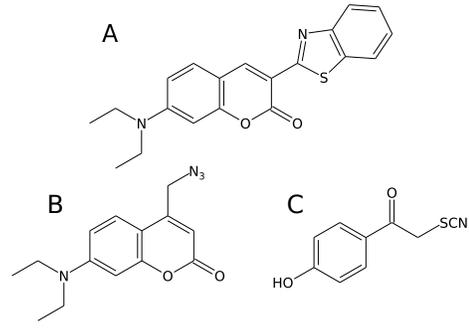}
    \caption{Structure of Coumarin 6 (\textbf{A}), {\color{black} DEACM-N$_3$, i.e.,} [7-(diethylamino)coumarin-4-yl]methyl-azide
             (\textbf{B}) {\color{black} and $p$HP-SCN,} i.e., \emph{para}-hydroxyphenacyl thiocyanate (\textbf{C}).}
\end{figure}

We apply both time-independent (TI) and time-dependent (TD) methods in conjunction with
a {\color{black} quadratic} vibronic model Hamiltonian that is parametrized in the full normal-mode
space, based upon ground-state and excited-state electronic structure
calculations. Besides displacements of the equilibrium geometry, the
Hamiltonian accurately represents Duschinsky rotation effects.

The TI approach uses a generalized version of the \emph{FCclasses}
method developed by one of
us\cite{santoro_effective_2007,santoro_effective_2007-1,santoro_effective_2008},
where a pre-screening
technique is adopted to select the relevant Franck-Condon (FC) overlap
integrals and obtain fully converged spectra even for high-dimensional molecular
systems. In the present context, this approach was specifically adapted such
as to include selective vibrational pre-excitation.

In a complementary fashion, TD methods are employed in order to
compute wavepacket autocorrelation functions, whose Fourier transforms yield
the vibronic absorption spectrum. Within the harmonic approximation and
including vibrational pre-excitation, autocorrelation functions are either
computed analytically or else using the Multi-Layer Multiconfiguration
Time-Dependent Hartree (ML-MCTDH) method\cite{wang_multilayer_2003,vendrell_multilayer_2011},
a recently developed variant of the MCTDH
method \cite{meyer_multi-configurational_1990,beck_multiconfiguration_2000,worth_using_2008}. In
the present context, all methods are expected to give the same results for
zero-temperature calculations including pre-excitation, but the efficiency of
these methods differs as a function of dimensionality and {\color{black} type of 
vibrational pre-excitation}.

The remainder of the manuscript is organized as follows. Sec.\ \ref{sec:theory}
details the TI and TD approaches employed
in this study and Sec.\ \ref{sec:analysis} introduces an analysis in terms of
spectral moments. Sec.\ \ref{sec:comp} summarizes the computational procedure, and
Sec.\ \ref{sec:results} presents results obtained for three representative
chromophores, together with experimental results for two of these systems.
Sec.\ \ref{sec:discussion} gives a discussion of the results and
Sec.\ \ref{sec:conclusion} concludes. Finally, several Appendixes add information
complementary to the main text. 

\section{Methods for the computation of vibrationally resolved electronic spectra from a
         vibrationally pre-excited state}\label{sec:theory}

Conventionally, the computation of vibrationally resolved electronic spectra proceeds from a
thermally averaged ensemble of excited vibrational states, populated according to a Boltzmann
distribution\cite{santoro_effective_2007-1}. In contrast, the present study is concerned with
vibronic excitation from a non-equilibrium state where a single vibrational normal mode is placed
into its first excited state by vibrational pre-excitation, while all other normal modes remain in
their respective ground state. {\color{black} Our study}
will be restricted to a
zero-temperature setting and we will focus on pre-excitations along high-frequency modes for which
thermal excitation is negligible. {\color{black}
  (However, Appendix C addresses finite-temperature spectra to assess whether
  thermal excitation significantly modifies the spectrum in the
  absence of pre-excitation.)} 

Vibrational states are defined in terms of the number of quanta in each vibrational normal mode:
$\ket{\bm{w}} = \ket{w_1} \otimes \ket{w_2} \otimes \cdots \otimes \ket{w_{N_{\text{vib}}}}$. A
vibrational state with pre-excitation of the $k$th mode {\color{black} in the electronic ground state ($\vert g \rangle$)}
is denoted by ${\color{black} \ket{\bm{w}_{gk}}  = \ket{0_{g1}} \otimes \ket{0_{g2}} \cdots \ket{1_{gk}} \cdots \ket{0_{gN_{\text{vib}}-1}}
\otimes \ket{0_{gN_{\text{vib}}}}}$ ${\color{black} = \ket{\bm{0}_g + 1_{gk}}}$. Combined vibrational-electronic (i.e.,
vibronic) states are denoted $\ket{\psi} = {\color{black} \ket{\bm{w}_n}} \otimes \ket{n}$ where the electronic space
$\{ \ket{n} \}$ is restricted to the electronic ground state ($\ket{g}$) and an
excited state ($\ket{e}$) in the following discussion. 

We are going to work in a normal-mode representation throughout, taking into account that
the normal modes of the initial state {\color{black} $\bm{Q}_g$} and the final state
{\color{black} $\bm{Q}_e$} are
different. Neglecting the effects of rotation {\color{black} (which can be minimized as explained, e.g., in Ref.\ \citenum{santoro_effective_2007})}, these sets of modes {\color{black} are} related by a linear
transformation as described by Duschinsky\cite{duschinsky37}:
\begin{equation}\label{eq:dusch}
    {\color{black} \bm{Q}_g = \bm{J} \bm{Q}_e + \bm{K}}
\end{equation}
where the transformation matrix $\bm{J}$ and the displacement vector $\bm{K}$ are defined by
\begin{equation}\label{eq:trafodef}
{\color{black}     \bm{J} = \bm{L}_g^{-1} \bm{L}_e \quad \text{and} \quad
    \bm{K} = \bm{L}_g^{-1} (\bm{q}_e^{\text{eq}} - \bm{q}_g^{\text{eq}})}
\end{equation}
with $\bm{L}$ the normal-mode matrix relating the normal modes $\bm{Q}$ to mass-weighted Cartesian
coordinates $\bm{q} = (q_1, q_2, ..., q_{3N})$,
\begin{equation}\label{eq:L_matrix}
    \bm{Q} = \bm{L}^{-1} (\bm{q} - \bm{q}^{\text{eq}})
\end{equation}
{\color{black} In \cref{eq:trafodef}, the equilibrium structures of the initial and final
state are termed $\bm{q}_g^{\text{eq}}$ and $\bm{q}_e^{\text{eq}}$}. 

In the following, the time-independent and time-dependent approaches will be detailed.

\subsection{Time-independent methods}

%$\frac{(2\pi)^2\omega N_A}{3\times1000\times\ln(10)\hbar c_0(4\pi\varepsilon_0)}$ (TI)
%$\frac{(2\pi)\omega N_A}{3\times1000\times\ln(10)\hbar c_0(4\pi\varepsilon_0)}$ (TD)

In the time-independent picture, the complete spectrum can be thought of as a weighted
superposition of vibronic (combined electronic and vibrational) transitions. This method is also
known as the sum-over-states method. The absorption spectrum {\color{black} resulting
  from excitation from the} ground state $\ket{g}$
to the excited state $\ket{e}$ can then be described as follows {\color{black} in terms of the
  extinction coefficient $\epsilon(\omega)$},\cite{santoro_effective_2007,tannor_tdqm}
{\color{black} 
\begin{eqnarray}\label{eq:spec_TI}
%    \sigma_{\text{abs}}(\omega) = \frac{4 \pi^2 \omega}{3c}
%   \sum_{\bm{w}_e} |\mu_{\bm{w}_g, \bm{w}_e}|^2 \, \delta(E_{\bm{w}_g} - E_{\bm{w}_e}  + \hbar \omega)
%     \epsilon(\omega) & = & \frac{(2 \pi)^2 N_A \omega}{3\, {\rm log}(10) \hbar c(4 \pi \epsilon_0)}
     \epsilon(\omega) & = & A\, \omega\,
     \sum_{\bm{w}_e} |\mu_{\bm{w}_g, \bm{w}_e}|^2 \, \delta\biggl(\frac{E_{\bm{w}_g} - E_{\bm{w}_e} - E_{\rm ad}}{\hbar} + \omega\biggr)
     \nonumber \\
\end{eqnarray}
}
{\color{black} where the prefactor is given as $A = {(2 \pi)^2 N_A}/{(3\, {\rm log}(10) \hbar c(4 \pi \epsilon_0))}$, with
$N_A$ the Avogadro constant. Further, 
$E_{\bm{w}_g}$ and $E_{\bm{w}_e}$} are the energies of the initial and final {\color{black} vibrational} states,
{\color{black} $E_{\text{ad}}$ is the zeroth-order (adiabatic) energy difference between the minima of the 
  two electronic states, and 
$\mu_{\bm{w}_g, \bm{w}_e} = \bra{\bm{w}_g}\braket{e | \hat{\mu} | g}\ket{\bm{w}_e} =
\braket{\bm{w}_g | \mu_{eg} (\bm{Q})| \bm{w}_e}$} is the transition dipole moment between the states.
{\color{black} In practice, one usually expresses $\epsilon(\omega)$ in units
  of [M$^{-1}$cm$^{-1}$] where M refers to the molar concentration (1 M = 1 mol dm$^{-3}$). The prefactor $A$
  then takes the numerical value $A$ = 703.3, if
all remaining terms are specified in atomic units.\cite{santoro_effective_2007}} 

Here, we will work within the Condon approximation, where the electronic transition dipole moment
is assumed to be coordinate-independent, $\mu_{eg} = \mu_0 = const$. In this case
{\color{black} $\mu_{\bm{w}_g, \bm{w}_e} = \mu_0 \braket{\bm{w}_g | \bm{w}_e}$ where $\braket{\bm{w}_g | \bm{w}_e}$} is the
multidimensional FC overlap integral between the initial and final vibrational states.

As a result, the time-independent framework for the computation of Eq.\ (\ref{eq:spec_TI}) focuses upon
the calculation of the FC overlap integrals {\color{black} $\braket{\bm{w}_g | \bm{w}_e}$} between
the initial and final vibrational states. In practice, it turns out that 
even in medium-sized molecules, the number of possible
final vibrational states {\color{black} $\ket{\bm{w}_e}$} is so large that the calculation of all FC integrals is
not feasible and a suitable subset must be selected. Here, we follow a scheme proposed by one of
us\cite{santoro_effective_2007} where the vibrational states are partitioned into so-called
{classes} $C_n$ where $n$ is the number of simultaneously excited normal modes in the final
state {\color{black} $\ket{\bm{w}_e}$} (\emph{FCclasses} method\cite{fcclassesprog}).

Naturally, the number of states in each class grows rapidly with increasing
$n$. For a spectrum at $\mathrm{T} = \SI{0}{\kelvin}$, all integrals, up to sufficiently high
maximum quantum numbers, are computed for the first two classes $C_1$ and $C_2$. Using these data,
an iterative procedure determines the best set of quantum numbers for class $C_n$ (up to $C_7$)
under the constraint that the total number of integrals to be computed for each class does not
exceed a pre-set maximum number $N_{\text{max}}$. By increasing $N_{\text{max}}$ the method can be
converged to arbitrary accuracy.

Spectra from initial states that are vibrationally excited
({\color{black} $\ket{\bm{w}_{gk}}$}) are computed adopting a modified strategy developed for thermally excited
states\cite{santoro_effective_2007-1}. In these cases, both the $C_1$ and $C_2$ transitions from
the ground state ({\color{black} $\ket{\bm{0}_g}$}) and the excited state {\color{black} ($\ket{\bm{w}_{gk}}$)} are computed and used
to select the relevant transitions of higher classes. As one would intuitively expect, excitations
of the final-state modes that are most similar to those excited in the initial state are
particularly important to reach reasonable convergence of the spectra. Therefore the subset of
final-state modes that project upon the
initial-state excited vibrational modes is determined and, for high classes $C_n$
($n > n_{\text{max}}$), final vibrational states where these modes are not excited are neglected.
Usually $n_{\text{max}} = 5$ provides a good compromise between accuracy and computational cost.

\subsection{Time-dependent methods}

The time-independent expression for the spectrum Eq.\ (\ref{eq:spec_TI}) can be reformulated in a
time-dependent framework as the Fourier transform of an autocorrelation
function, {\color{black} such that Eq.\ (\ref{eq:spec_TI}) takes the alternative form:}\cite{tdspec_book,tannor_tdqm}
\begin{equation}\label{eq:spec_TD}
%    \sigma_{\text{abs}} (\omega) = \frac{2 \pi \omega}{3 c}
   {\color{black} \epsilon(\omega) = \frac{A\, \omega}{2\pi}}
    \int \chi (t, T) e^{\mathrm{i}(\omega - E_{\text{ad}} / \hbar)t} dt
\end{equation}
%where $E_{\text{ad}}$ is the zeroth-order (adiabatic) energy difference between the
%two states' minima.
In general, the autocorrelation function $\chi (t, T)$ can be
expressed as follows\cite{tdspec_book,avila_ferrer_first-principle_2014,tannor_tdqm}:
\begin{equation}\label{eq:thermauto}
    \chi (t, T) = \text{Tr} \left[ \hat{{\mu}}_{eg}\ e^{-\mathrm{i} \hat{H}_e t / \hbar}
    \hat{{\mu}}_{eg}\ e^{\mathrm{i} \hat{H}_g t / \hbar} \hat{\rho}_g (0) \right]
\end{equation}
where Tr denotes the trace operation, and $\hat{\rho}_g(0) $ is the combined
vibrational and electronic density operator referring to the initial state
($g$), i.e., $\hat{\rho}_g(0) = \hat{\rho}_g^{\text{vib}}(0) \otimes \ket{g}\! \bra{g}$. The dipole
operator is again taken to be coordinate
independent, $\hat{{\mu}}_{eg} = \mu_0 ( \ket{e}\! \bra{g} +\text{h.c.})$, and $\hat{H}_g$ and
$\hat{H}_e$ are the Hamiltonians of the ground state and the
excited state, respectively.

General expressions for the above autocorrelation function starting from a
thermally populated initial state have been discussed
elsewhere\cite{tdspec_book,avila_ferrer_first-principle_2014,baiardi_general_2013,Pollak2004,Pollak2008,Lin2003,YanMukamel86,Mukamel85,Kubo55},
and the autocorrelation function for a \SI{0}{\kelvin} state has been derived
in Ref.\ [\onlinecite{baiardi_general_2013}]. Here, we focus specifically upon
the case of an initial vibrationally pre-excited state at \SI{0}{\kelvin},
{\color{black}
\begin{equation}
    \hat{\rho}_g(0) = \ket{\bm{0}_g + 1_{gk}} \! \bra{\bm{0}_g + 1_{gk}}
\end{equation}}
leading to the following form of the autocorrelation function,
{\color{black}
\begin{eqnarray}\label{eq:chik}
    \chi (t, T=0) \equiv \chi_k (t) =\, &&\mu_0^2\, e^{\mathrm{i}(E_0 / \hbar + \omega_{gk})t}
    \nonumber\\ &&\Braket{\bm{0}_g + 1_{gk} | e^{-\mathrm{i} \hat{H}_{e} t} | \bm{0}_g + 1_{gk}}
\end{eqnarray}
}
where $E_0$ is the vibrational zero-point energy of the initial state and $\omega_{gk}$ is the
angular frequency of the pre-excited mode.

In the following, two approaches of obtaining the correlation function
are introduced, i.e., by evaluation of analytical expressions and by numerical wavepacket
propagation, respectively. Within the harmonic approximation, both approaches give the same result,
{\color{black} apart from numerical inaccuracies.}
In situations where anharmonicity or non-adiabatic couplings
play an important role, it is mandatory, though, to switch to more general technique of wavepacket
propagation, making no assumptions about the shape of the potential energy surfaces (PESs).

{\color{black} The analytical approach detailed below bears some relation to the developments of Refs.\
  [\onlinecite{huh_coherent_2012,baiardi_general_2013,Barone2014}] where general time-dependent formulations were addressed that
  also include excitation of selected modes.} 

\subsubsection{Analytical expression for the autocorrelation function}\label{sec:auto}

%Here we will focus on a state where a specific vibrational mode
%has been pre-excited while all other modes are in their respective ground state. In the following,
%the derivation will be briefly outlined. The complete derivation is given in the Appendix. We will
%use the subscripts $i$ and $f$ to refer to the initial and final electronic states, respectively.

%In the case of an initial state where one quantum has been put into mode $k$, the autocorrelation
%function \cref{eq:thermauto} can be written as

To obtain an analytical expression for $\chi_k (t)$ of Eq.\ (\ref{eq:chik}), we employ the
normal-mode representation, in line with the treatment of Refs.\
[\onlinecite{avila_ferrer_first-principle_2014,baiardi_general_2013,peng_vibration_2010}].
Since the initial state normal modes $\bm{Q}_g$ form a
complete basis set, the coordinate representation of $\chi_k (t)$ can be written as follows by
inserting the identity twice,
\begin{eqnarray}\label{eq:corrcor}
    \chi_k (t) = \mu_0^2\, e^{\mathrm{i}(E_0 / \hbar + \omega_{gk})t} \int d\bar{\bm{Q}}_g
    \int d\bm{Q}_g &&{\color{black} \braket{\bm{0}_g + 1_{gk} | \bm{Q}_g} }\nonumber\\
                 \braket{\bm{Q}_g | e^{-\mathrm{i} \hat{H}_e t} | \bar{\bm{Q}}_g}
                 &&{\color{black} \braket{\bar{\bm{Q}}_g | \bm{0}_g + 1_{gk}}}
                 \nonumber \\
\end{eqnarray}
The coordinate representation of the vibrationally pre-excited initial state is given as follows,
\begin{equation}\label{eq:HO_1st_exc_state}
   {\color{black} 
  \braket{\bm{Q}_g | \bm{0}_g + 1_{gk}} =  Q_k \sqrt{2 \Gamma_{gk}}
    \braket{\bm{Q}_g | \bm{0}_g }
    }
\end{equation}
with the harmonic oscillator ground state 
\begin{equation}\label{eq:HO_ground_state}
    {\color{black} \braket{\bm{Q}_g  | \bm{0}_g }} = \frac{\det(\bm{\Gamma}_g)^{1/4}}{\pi^{N/4}}
    \exp \left[- \frac{\bm{Q}_g^T \bm{\Gamma}_g \bm{Q}_g}{2} \right]
\end{equation}
where $\bm{\Gamma}_g$ is a diagonal matrix containing the reduced frequencies of the initial state
($g$) normal modes $(\bm{\Gamma}_g)_{kk} = \Gamma_{gk} = \omega_{gk} / \hbar$. 
>From \cref{eq:HO_1st_exc_state,eq:HO_ground_state}, we note that
{\color{black} $\braket{\bm{Q}_g | \bm{0}_g + 1_{gk}} = \braket{\bm{0}_g + 1_{gk} | \bm{Q}_g}$}. 
Inserting the latter expression into Eq.\ (\ref{eq:corrcor}) results in
\begin{widetext}
\begin{equation}\label{eq:corr_initq}
    \chi_k (t) = 2 \mu_0^2 \Gamma_{gk} e^{\mathrm{i}(E_0 / \hbar + \omega_{gk})t}
    \frac{\det(\bm{\Gamma}_g)^{1/2}}{\pi^{N/2}} \int\!\! d\bar{\bm{Q}}_g \!\int\!\! d\bm{Q}_g
    Q_{gk} \bar{Q}_{gk} \exp \left[- \frac{\bm{Q}_g^T \bm{\Gamma}_g \bm{Q}_g}{2} \right]
    \braket{\bm{Q}_g | e^{-\mathrm{i} H_e t} | \bar{\bm{Q}}_g}
    \exp \left[- \frac{\bar{\bm{Q}}_g^T \bm{\Gamma}_g \bar{\bm{Q}}_g}{2} \right].
\end{equation}
In order to represent the matrix element of the final state propagator,
$\braket{\bm{Q}_g | e^{-\mathrm{i} \hat{H}_e t} | \bar{\bm{Q}}_g}$ in terms of the final state
normal modes, we now introduce two complete sets of final state coordinates $\bm{Q}_e$:
\begin{eqnarray}\label{eq:ap_corr1}
    \chi_k (t) = 2 \mu_0^2 \Gamma_{gk} e^{\mathrm{i}(E_0 / \hbar + \omega_{gk})t}
    &&\frac{\det(\bm{\Gamma}_g)^{1/2}}{\pi^{N/2}}
    \int d\bar{\bm{Q}}_g \int d\bm{Q}_g \int d\bar{\bm{Q}}_e \int d\bm{Q}_e \nonumber \\
    &&Q_{gk} \bar{Q}_{gk} \exp \left[- \frac{\bm{Q}_g^T \bm{\Gamma}_g \bm{Q}_g}{2} \right]
    \braket{\bm{Q}_g | \bm{Q}_e}
    \braket{\bm{Q}_e | e^{-\mathrm{i} \hat{H}_e t} | \bar{\bm{Q}}_e}
    \braket{\bar{\bm{Q}}_e | \bar{\bm{Q}}_g}
    \exp \left[- \frac{\bar{\bm{Q}}_g^T \bm{\Gamma}_g \bar{\bm{Q}}_g}{2} \right].
\end{eqnarray}
We can now use Feynman's path integral expression to evaluate the matrix element of the propagator
\begin{equation}\label{eq:ap_feynmann}
      \braket{\bm{Q}_e | e^{-\mathrm{i} \hat{H}_e t} | \bar{\bm{Q}}_e}
    = \sqrt{\frac{\det(\bm{a}_e(t))}{(2 \pi \mathrm{i} \hbar)^N}}
      \exp \left\lbrace \frac{\mathrm{i}}{\hbar}
      \left[ \frac{1}{2} \bm{Q}_e^T \bm{b}_e(t) \bm{Q}_e
      + \frac{1}{2} \bar{\bm{Q}}_e^T \bm{b}_e(t) \bar{\bm{Q}}_e
      - \bm{Q}_e^T \bm{a}_e(t) \bar{\bm{Q}}_e \right] \right\rbrace,
\end{equation}
where $\bm{a}_e(t)$ and $\bm{b}_e(t)$ are diagonal matrices with
\begin{equation}
    ( \bm{a}_e )_{kk}(t) \equiv a_{ek} (t) = \frac{\Gamma_{ek}}{\sin (\hbar \Gamma_{ek} t)}
    \quad \text{and} \quad
    ( \bm{b}_e )_{kk}(t) \equiv b_{ek} (t) = \frac{\Gamma_{ek}}{\tan (\hbar \Gamma_{ek} t)}.
\end{equation}
Since $\bm{Q}_g$ and $\bm{Q}_e$ are orthonormal, the overlap is given as $\braket{\bm{Q}_g | \bm{Q}_e}
= \delta(\bm{Q}_g - \bm{J} \bm{Q}_e - \bm{K})$. Inserting these relations into Eq.\ (\ref{eq:ap_corr1}) results in the Gaussian integral
\begin{multline}
    \chi_k (t) = 2 \mu_0^2 \Gamma_{gk} e^{\mathrm{i}(E_0 / \hbar + \omega_{gk})t}
    \sqrt{\frac{\det(\bm{\Gamma}_g) \det(\bm{a}_e(t))}{\pi^N (2 \pi \mathrm{i} \hbar)^N}}
    \exp \left[-\bm{K}^T \bm{\Gamma}_g \bm{K} \right] \int d\bar{\bm{Q}}_e \int d\bm{Q}_e
\\  \left( K_k^2 + K_k \sum_l J_{kl} \left(Q_{el} + \bar{Q}_{el} \right)
    + \sum_l \sum_m J_{kl} J_{km} Q_{el} \bar{Q}_{em} \right)
\\  \exp \left\lbrace \frac{\mathrm{i}}{\hbar} \left[ \mathrm{i}\hbar \bm{K}^T \bm{\Gamma}_g \bm{J}
    (\bm{Q}_e + \bar{\bm{Q}}_e) + \frac{1}{2} \bm{Q}_e^T \bm{B}(t) \bm{Q}_e
    + \frac{1}{2} \bar{\bm{Q}}_e^T \bm{B}(t) \bar{\bm{Q}}_e - \bm{Q}_e^T \bm{a}_e(t) \bar{\bm{Q}}_e
    \right] \right\rbrace
    \label{eq:Gauss_integral_intermediate}
\end{multline}
\end{widetext}
where we introduced the matrix
\begin{equation}
    \bm{B}(t) = \mathrm{i}\hbar \bm{J}^T \bm{\Gamma}_g \bm{J} + \bm{b}_e(t)
\end{equation} 
As detailed in Appendix \ref{sec:appA}, this expression can be integrated analytically and is
further recast as follows in a compact and transparent form,
\begin{eqnarray}\label{eq:chik_final}
    &&\chi_k(t) = 2 \Gamma_{gk} e^{\mathrm{i}\hbar\Gamma_{gk} t} \chi_{FC}^0 (t)\nonumber \\
    &&\left( K_k^2 + 2 \sum_i K_k J_{ki} \tilde{\bm{D}}_i (t)
    + \sum_{ij} J_{ki} J_{kj} \tilde{\bm{A}}_{ij} (t) ) \right)
\end{eqnarray} 
where $\chi^0_{FC}(t)$ is the autocorrelation function at
\SI{0}{\kelvin} \cite{baiardi_general_2013} in the absence of initial vibrational excitation,
\begin{eqnarray}
    \chi^0_{FC}(t) = \mu_0^2 &&\sqrt{\frac{\det(\bm{a}'_g(t)\bm{a}_e(t))}
    {(\mathrm{i} \hbar)^{2N} \det(\bm{CD})}} \nonumber \\
    &&\exp \left[-\bm{K}^T \bm{\Gamma}_g \bm{K} + \bm{\lambda}^T \bm{D}^{-1} \bm{\lambda} \right],
\end{eqnarray}
with $\bm{\lambda}^T = \bm{K}^T \bm{\Gamma}_g \bm{J}$. Further, the following auxiliary
quantities were introduced,
\begin{eqnarray}\label{eq:DAtilde}
    \tilde{\bm{D}}(t) & = & -\bm{D}^{-1}(t) \bm{J}^T \bm{\Gamma}_g \bm{K} \nonumber \\
    \tilde{\bm{A}}(t) & = & \tilde{\bm{D}}(t) \tilde{\bm{D}}^T(t)
    + \frac{1}{2} \left( \bm{D}^{-1}(t) - \bm{C}^{-1}(t) \right)
\end{eqnarray}
and, in turn,
\begin{eqnarray}\label{eq:CD}
    \bm{C}(t) & = & -\frac{i}{\hbar} ( \bm{B}(t) + \bm{a}_e(t) ) \nonumber \\
    \bm{D}(t) & = & -\frac{i}{\hbar} ( \bm{B}(t) - \bm{a}_e(t) )
\end{eqnarray}
%where ${\bm{B}}(t) = i \hbar \bm{J}^T \bm{\Gamma}_g \bm{J} + \bm{b}_e(t)$. 
The above expressions for $\tilde{\bm{D}}$ and $\tilde{\bm{A}}$ and the overall expression for the
correlation function are very similar to expressions that were previously obtained to describe
Herzberg-Teller transitions, where the transition dipole depends linearly on the
coordinates\cite{avila_ferrer_first-principle_2014,baiardi_general_2013,peng_vibration_2010,Peluso2012}. 

\subsubsection{Autocorrelation function by wavepacket propagation}\label{sec:wppropa}

The autocorrelation function of Eq.\ (\ref{eq:chik}) can be obtained alternatively by propagating
a wavepacket corresponding to the initial vibrational state
{\color{black} $\ket{\bm{0}_g + 1_{gk}}$} evolving on the
excited-state PES,
{\color{black} 
\begin{equation}
  \ket{\psi_e(0)} =  \ket{{\bm{0}}_g + 1_{gk}} \otimes \ket{e}
\end{equation}
}
such that\cite{tannor_tdqm}  
\begin{eqnarray}\label{chik}
    \chi_k (t) & = & \mu_0^2\, e^{\mathrm{i}(E_0 / \hbar + \omega_{gk})t}
    \Braket{ \psi_e(0) | e^{-\mathrm{i} \hat{H}_{e} t} | \psi_e(0) } \nonumber \\
    & = & \mu_0^2\, e^{\mathrm{i}(E_0 / \hbar + \omega_{gk})t} \Braket{ \psi_e(0) | \psi_e(t) }
\end{eqnarray}
In this approach, it is convenient to work in the normal-mode representation of the electronic
ground state ($g$), where the initial wavefunction is separable with respect to the normal-mode
eigenfunctions, one of which corresponds to a vibrationally excited state,
\begin{equation}
    \ket{\psi_e (\bm{Q}_g, 0)} = \left( \varphi_{k1}({Q}_{gk})
    \prod_{n\neq k} \varphi_{n0}({Q}_{gn}) \right) \otimes \ket{e}
\end{equation}
Propagation necessitates a representation of the excited-state PES in terms of ground-state normal
modes, which is obtained as follows, using the transformation \cref{eq:dusch}. 
Given that the excited-state PES is diagonal in terms of its normal modes:
{\color{black} 
\begin{equation}\label{eq:Vfinal}
    V_e (\bm{Q}_e) = \frac{1}{2} \bm{Q}_e^T \bm{F} \bm{Q}_e + E_{\text{ad}}
\end{equation}
where $\bm{F}$ is the diagonal matrix of the final state normal modes' force constants, we can 
invert \cref{eq:dusch} to yield
\begin{equation}
    \bm{Q}_e = \widetilde{\bm{J}} \bm{Q}_g + \widetilde{\bm{K}} \quad
    \text{with} \quad \widetilde{\bm{J}} = \bm{J}^T \quad
    \text{and} \quad \widetilde{\bm{K}} = -\bm{J}^T \bm{K}.
\end{equation}
Inserting this expression for $\bm{Q}_e$ into \cref{eq:Vfinal}, we obtain
\begin{eqnarray}
    V_e (\bm{Q}_g) &=& \frac{1}{2} \left( \widetilde{\bm{J}} \bm{Q}_g + \widetilde{\bm{K}} \right)^T
    \bm{F} \left( \widetilde{\bm{J}} \bm{Q}_g + \widetilde{\bm{K}} \right) + E_{\text{ad}}
    \nonumber \\
    &=& \frac{1}{2} \bm{Q}_g^T \widetilde{\bm{J}}^T \bm{F} \widetilde{\bm{J}} \bm{Q}_g
    + \widetilde{\bm{K}}^T \bm{F} \widetilde{\bm{J}} \bm{Q}_g
    + \frac{1}{2} \widetilde{\bm{K}}^T \bm{F} \widetilde{\bm{K}} \nonumber \\ &&+ E_{\text{ad}}
\end{eqnarray}
and going back to the original $\bm{J}$ and $\bm{K}$, the final expression for the PES reads
\begin{eqnarray}\label{eq:mctdhpes}
    V_e (\bm{Q}_g) = \frac{1}{2} \bm{Q}_g^T \bm{J} \bm{F} \bm{J}^T &&\bm{Q}_g
                  - \bm{K}^T \bm{J} \bm{F} \bm{J}^T \bm{Q}_g
\nonumber \\    &&+ \frac{1}{2} \bm{K}^T \bm{J} \bm{F} \bm{J}^T \bm{K} + E_{\text{ad}}
\end{eqnarray}
}
This expression for the excited-state PES has been employed in conjunction with the
ML-MCTDH wavepacket propagation method, as further detailed below. 

\section{Analysis of computed spectra}\label{sec:analysis}

In order to quantitatively characterize the influence of vibrational pre-excitation on the
electronic absorption spectrum, it is useful to analyze its influence on the first and second
spectral moments. The first moment is equivalent to the {expectation value} or the center of
gravity of the spectrum, whereas the second moment corresponds to the width. In general, the $N$th
moment of a continuous  distribution function is given by
\begin{equation}
    \mathcal{M}^{(N)} = \int\limits_{-\infty}^{\infty} x^N f(x) dx.
\end{equation}
For the spectrum computed at the FC level within the harmonic approximation, the first and second
moments can be calculated analytically\cite{tispec_book}. 
The first moment of a spectrum starting
from the vibrational ground state
{\color{black} $\ket{\bm{0}_g}$} state is given by
\begin{equation}\label{eq:mom10}
    \mathcal{M}^{(1)}_0 = E_{\text{ad}} + \frac{1}{2} \bm{K}^T \widetilde{\bm{F}} \bm{K}
    + {\color{black} \frac{1}{4} \sum_i \frac{\widetilde{F}_{ii} - \Omega_{gi}^2}{\Omega_{gi}},}
\end{equation}
where {\color{black} $\Omega_{gi}$}
is the frequency of the $i$th ground state normal mode and
$\widetilde{\bm{F}} = \bm{J} \bm{F} \bm{J}^T$ is the matrix of the excited state force constants
projected onto the ground state normal modes (cf. Eq.\ (\ref{eq:mctdhpes})). The second moment
reads as follows (see Ref.\ [\onlinecite{tispec_book}] for a more general expression including temperature
effects),
{\color{black} 
\begin{eqnarray}
    \mathcal{M}^{(2)}_0 &&=
    E_{\text{v}}^2
    + \frac{E_{\text{v}}}{2} \sum_i \frac{\widetilde{F}_{ii} - \Omega_{gi}^2}{\Omega_{gi}} \nonumber
\\    &&+ \frac{1}{8} \sum_i \sum_{j > i}
      \frac{(\widetilde{F}_{ii} - \Omega_{gi}^2) (\widetilde{F}_{jj} - \Omega_{gj}^2)}{\Omega_{gi} \Omega_{gj}}
    + \frac{1}{2} \sum_i \frac{g_i^2}{\Omega_{gi}} \nonumber
\\    &&+ \frac{3}{8} \sum_i \frac{(\widetilde{F}_{ii} - \Omega_{gi}^2)^2}{\Omega_{gi}^2}
    + \frac{1}{2} \sum_i \sum_{j > i} \frac{\widetilde{F}_{ij}^2}{\Omega_{gi} \Omega_{gj}}
\end{eqnarray}
}
where $E_{\text{v}}$ is the vertical transition energy at the ground state minimum, corresponding
to the first two terms in \cref{eq:mom10} and $\bm{g} = -\bm{K}^T \widetilde{\bm{F}}$ is the
gradient on the PES of the excited state along the ground state normal modes. If we start from a
pre-excited state {\color{black} $\ket{\bm{0}_g + 1_{gk}}$},
the first moment is given by
{\color{black}
\begin{equation}
    \mathcal{M}^{(1)}_k = \mathcal{M}^{(1)}_0
    + \frac{1}{2} \frac{\widetilde{F}_{kk} - \Omega_{gk}^2}{\Omega_{gk}},
\end{equation}
while the second moment becomes
\begin{eqnarray}
    \mathcal{M}^{(2)}_k &&= \mathcal{M}^{(2)}_0
    + E_{\text{v}} \frac{\widetilde{F}_{kk} - \Omega_{gk}^2}{\Omega_{gk}} \nonumber \\
    &&+ \frac{\widetilde{F}_{kk} - \Omega_{gk}^2}{2 \Omega_{gk}}
      \sum_{ i \neq k} \frac{\widetilde{F}_{ii} - \Omega_{gi}^2}{\Omega_{gi}}
    + \frac{g_k^2}{\Omega_{gk}} \nonumber \\
    &&+ \frac{15}{8} \frac{(\widetilde{F}_{kk} - \Omega_{gk}^2)^2}{\Omega_{gk}^2}
    + \sum_{ i \neq k} \frac{2 \widetilde{F}_{ik}}{\Omega_{gi} \Omega_{gk}}.
\end{eqnarray}
}
From the first and second moments, the standard deviation of the spectrum can be obtained:
\begin{equation}
    \sigma = \left( \mathcal{M}^{(2)} - \left(\mathcal{M}^{(1)}\right)^2 \right)^{1/2}.
\end{equation}

Furthermore, to get an \emph{a priori} estimate of the suitability of a particular normal mode for
VIPER excitation, we compute the ratio of the intensities of the transition from the pre-excited
vibrational state
{\color{black} $\ket{\bm{0}_g + 1_{gk}}$} versus the ground vibrational state {\color{black} $\ket{\bm{0}_g}$} to the
ground vibrational state of the excited electronic state
{\color{black} $\ket{\bm{0}_e}$}:
{\color{black}
\begin{equation}
    \frac{\Braket{\bm{0}_g + 1_{gk} | \bm{0}_e}}{\Braket{\bm{0}_g | \bm{0}_e}} = \frac{S_k}{\sqrt{2}}
\end{equation}
with the column vector $\bm{S}$ given by
{\color{black} 
\begin{equation}\label{eq:bvec}
    \bm{S} = 2 \bm{\delta}_g
    \left( \bm{1} - \bm{\Omega}_g^{1/2} \bm{J} \bm{X}^{-1} \bm{J}^T \bm{\Omega}_g^{1/2} \right),
\end{equation}
with
\begin{equation}
    \bm{X} = \bm{J}^T \bm{\Omega}_g\bm{J} + \bm{\Omega}_e.
\end{equation}
}}
where
{\color{black} $\bm{\Omega}_g$ and $\bm{\Omega}_e$} are the diagonal matrices of the ground and excited state's
vibrational frequencies, respectively. Further,
\begin{equation}
{\color{black} \bm{\delta} = \bm{K}^T \bm{\Omega}_g^{1/2} }
\end{equation}
is the
vector of the {dimensionless displacements} along the ground state normal modes. These
displacements, {\color{black} which are related to the Huang-Rhys factors\cite{Mukamel}}, are a useful quantity for determining the difference between the two states'
equilibrium structures.

It is evident from \cref{eq:bvec} that the ratio of the vibronic transitions in
question is mainly proportional to the dimensionless shift, i.e. the displacement between the
equilibrium structures. The Duschinsky mixing and the vibrational modes' frequency change upon
excitation also enters, though, in the second term, i.e., the matrix expression in \cref{eq:bvec}.

{\color{black} When calculating the first and second spectral moments, along with the
  quantities $\bm{S}$ and $\bm{\delta}$,  
  exact results are obtained -- by definition -- from the analytical TD approach
  while the TI approach may show inaccuracies due to the truncation of the total number of
  FC interals, as discussed above. The ML-MCTDH approach, in turn, may also exhibit
  numerical convergence issues, as further illustrated below.} 

\section{Computational procedure}\label{sec:comp}

Three molecular systems were investigated in this work, two of which have already been studied
experimentally\cite{bredenbeck_viper2014_int,kern_steering_inprep}. The laser dye Coumarin 6 is the
first system that was investigated experimentally using the VIPER
approach\cite{bredenbeck_viper2014_int}.
The other two systems, i.e., [7-(diethylamino)coumarin-4-yl]methyl-azide (DEACM-N$_3$) and
\emph{para}-Hydroxyphenacyl
thiocyanate (\emph{p}HP-SCN), are model systems for photo-cleavable caging
groups\cite{klan_photoremovable_2012}, with the azide and thiocyanate groups representing the
leaving group, respectively.

Electronic structure calculations were performed with the \emph{Gaussian09} package, revision
\emph{D.01}\cite{g09}, using density functional theory (DFT) for the ground state and its
time-dependent extension (TD-DFT) for the excited state. Geometry optimizations and harmonic
vibrational analyses in the ground state were performed using analytical first and second
derivatives, respectively. For the excited state, numerical differentiation of analytic gradients
was used to obtain second derivatives. Tight optimization criteria in combination with fine
integration grids (\texttt{int=ultrafine}) were used throughout. Solvent effects were treated using
the polarizable continuum model (PCM). For Coumarin 6 and DEACM-N$_3$ the long-range-corrected
hybrid functional \mbox{$\omega$B97x-D}\cite{chai_long-range_2008} was used exclusively. Due to
computational issues (see below), different density functionals were used for
\emph{p}HP-SCN, including the CAM-B3LYP\cite{yanai_new_2004} and PBE0\cite{adamo_toward_1999}
hybrid functionals. All electronic structure calculations were performed using the
Def2-TZVP\cite{weigend_balanced_2005} triple-zeta basis set.

{\color{black} The excited-state PES was constructed using the
adiabatic Hessian (AH) approach\cite{ferrer_comparison_2012}, expanding the PES around its
state-specific minimum.}

For the numerical wavepacket propagation, the multi-layer ML-MCTDH
variant\cite{wang_multilayer_2003} of the Multi-Configuration Time-Dependent Hartree
method\cite{beck_multiconfiguration_2000} was employed, using the
Heidelberg MCTDH package, version 8.5.5\cite{MLMctdhpackageBla}. The multi-layer tree
representing the wavefunction partitioning was
constructed from the bottom up by successively combining pairs of modes, as detailed in Appendix
\ref{sec:appB}. The primitive basis was built in a harmonic-oscillator Discrete Variable
Representation (DVR).

Calculations of vibrationally resolved absorption spectra using the TI formalism
were performed with a development version of the \emph{FCclasses}
code\cite{fcclassesprog}.
The TI computations were performed with $N_{\text{max}} = 10^8$, and the resulting stick spectra
were convoluted with a Lorentzian lineshape with a HWHM (Half Width at Half Maximum) of \SI{0.01}{\electronvolt}.

Complementary TD calculations were performed by the two methods described above, i.e., the
analytical approach of \cref{sec:auto} (also implemented in the \emph{FCclasses} development
version) and
the numerical wavepacket approach detailed in \cref{sec:wppropa}.
%All approaches refer to the same
%PES in the AH formulation as employed in the TI calculations.
For the numerical wavepacket
approach, the PES {\color{black} in the AH formulation} was expressed in ground-state normal modes, see Eq.\ (\ref{eq:mctdhpes}); for
this purpose, an in-house code was used. Time correlation functions obtained by either the
analytical or numerical TD approach were subsequently Fourier transformed\cite{NumericalRecipes}
to yield the absorption spectrum. Before Fourier transforming, the correlation functions
$\chi^0 (t)$ were multiplied by a damping term, 
\begin{equation}
    \chi (t) = \chi^0 (t) \cos \left( \frac{\pi}{2 t_{\text{max}}} t \right) e^{- |t| / \tau}
\end{equation}
where $t_{\text{max}}$ is the time up to which the correlation function is calculated and $\tau$
is the damping time. To match the spectral broadening applied to the spectra from the TI approach, 
a value of $\tau = \SI{65.8}{\femto\second}$ was chosen, corresponding to a Lorentzian
HWHM of \SI{0.01}{\electronvolt}.

\section{Results}\label{sec:results}

{\color{black} As pointed out in the Introduction, our procedure captures
  the VIPER excitation scheme depicted in Fig.\ 1, which combines a resonant
  IR pulse with a subsequent UV/Vis pulse. The VIPER pulse sequence as such is
  more complex, and includes selection of the pre-excited species by an
  off-resonant UV/Vis pulse and final detection by an IR probe
  pulse.\cite{vanwilderen_ultrafast_2015,bredenbeck_viper2014_int} In our
  approach, which is restricted to first-order spectroscopic quantities, we
  cannot quantitatively calculate VIPER signals, but we are able to predict
  which modes are most suitable for pre-excitation within the VIPER scheme.

VIPER active modes} 
should fulfill
several criteria: they should have a strong infrared absorption cross section to achieve the
largest possible signal strength and they should be well separated to allow for selective
excitation using an infrared pulse with an FWHM
{\color{black} (Full Width at Half Maximum)} of approximately \numrange{8}{20}
\si{\per\centi\metre}. For the systems investigated here, experiments have been carried out
in the range between \SI{1500}{\per\centi\metre} and \SI{1900}{\per\centi\metre}, which
turns out to contain several promising modes in terms of VIPER activity. 
For all three systems, the lowest singlet excited state with significant oscillator strength was
chosen for further investigation.

In the following, we discuss theoretical results
{\color{black} for electronic absorption spectra including vibrational
  pre-excitation} in
relation to selected experimental Fourier-Transform Infrared (FTIR) and VIPER results.

\begin{figure}
    \includegraphics[width=0.95\columnwidth]{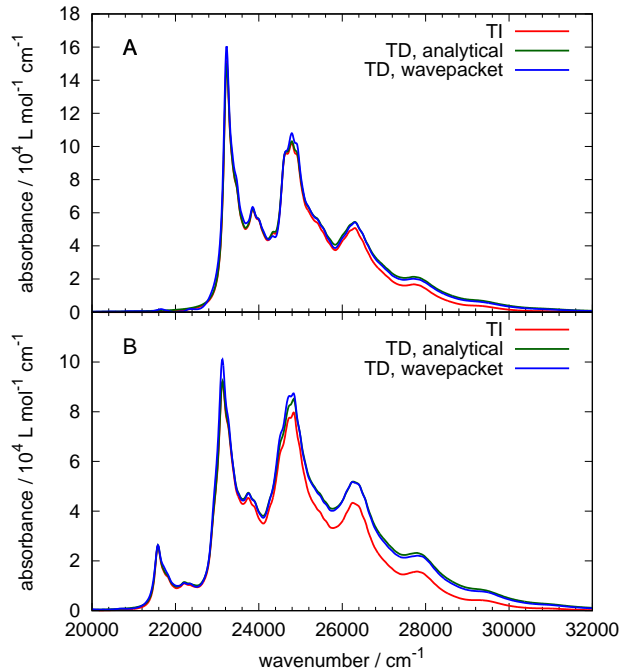}
    \caption{Computed absorption {\color{black} spectra} of Coumarin 6 in THF obtained from
             \mbox{$\omega$B97x-D}/Def2-TZVP calculations. {\color{black} (A) without
              vibrational pre-excitation, (B) with
             pre-excitation of the lower-frequency ring mode (see also Fig.\ 4A).} Spectra were computed using the
             TI approach (red), Fourier transformation of the analytical autocorrelation function
             (green) and of the wavepacket autocorrelation function (blue). The {\color{black} spectra} obtained
             from the wavepacket propagation have been red-shifted by \SI{90}{\per\centi\metre}
             (see text for discussion).}\label{fig:c6_0k}
\end{figure}

\begin{figure*}
    \includegraphics[width=0.9\textwidth]{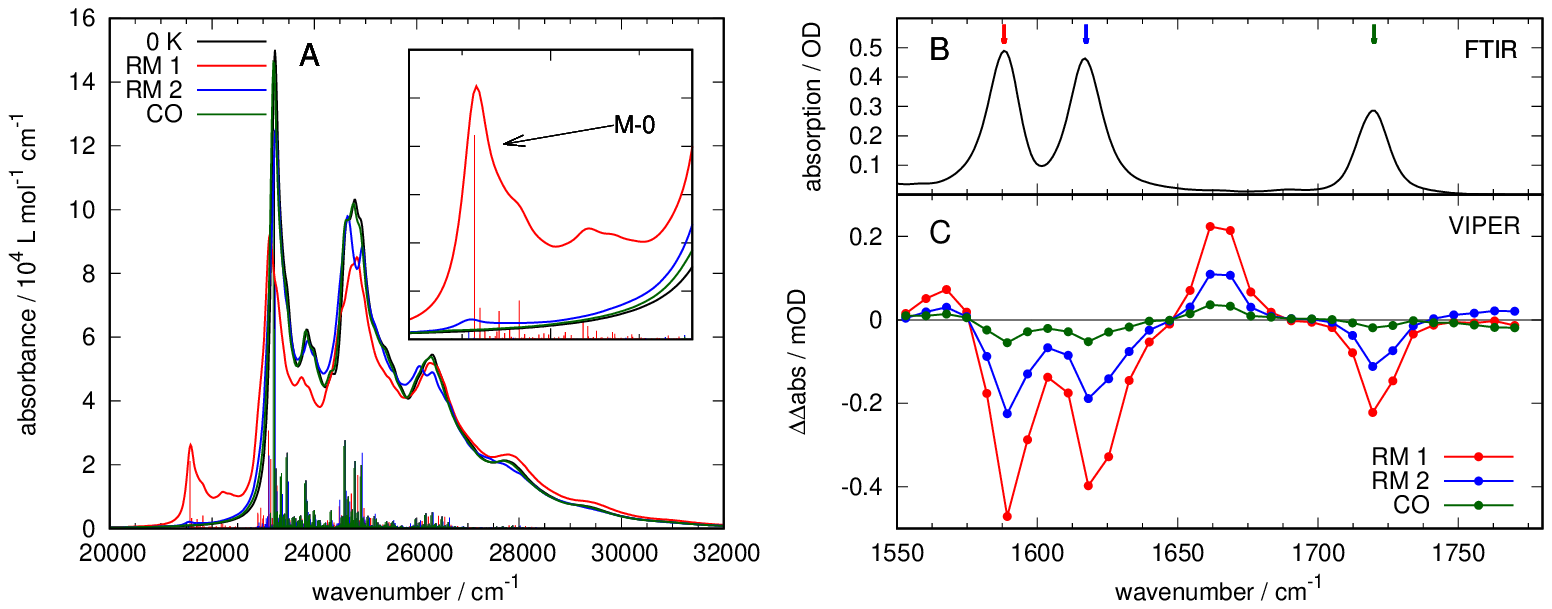}
    \caption{\textbf{(A)} Computed absorption spectra of Coumarin 6 in THF using the analytical
             TD approach, {\color{black} along with stick spectra obtained from the TI approach, based on a parametrization
             using \mbox{$\omega$B97x-D}/Def2-TZVP data.}
             Spectra are shown without vibrational pre-excitation (black), with pre-excitation of the
             lower-frequency ring mode (red), the higher-frequency ring mode (blue) and the CO
             stretch mode (green). \textbf{(B)} Experimental FTIR absorption spectrum of Coumarin 6
             in THF. \textbf{(C)} VIPER spectra of Coumarin 6 in THF after pre-excitation of the
             lower-frequency ring mode (red), the higher-frequency ring mode (blue) and the CO
             stretch mode (green). Experimental spectra were recorded using a concentration of
             10 mM and a layer thickness of 250 {$\mu$m}. {\color{black} Details about the
               experimental set-up can be found in Refs.\ \citenum{bredenbeck_viper2014_int,vanwilderen_ultrafast_2015,kern_steering_inprep}.}
}\label{fig:c6_exptheo}
\end{figure*}

\subsection{Coumarin 6}

Coumarin 6 contains 43 atoms resulting in 123 vibrational normal modes. The most interesting modes
in the context outlined above are the carbonyl stretch mode and two ring distortion modes in the
coumarin moiety. Coumarin 6 shows a bright HOMO-LUMO transition with
$\pi$-$\pi^*$ character on the coumarin moiety, exhibiting large oscillator strength.

\Cref{fig:c6_0k} shows {\color{black} computed absorption spectra} of Coumarin 6 at zero temperature in
THF, {\color{black} both without vibrational pre-excitation (panel (A)) and with vibrational pre-excitation
  of the lower-frequency ring mode (panel (B)). Good agreement is obtained between the
  different methods outlined above. In particular, the analytical TD results and the MCTDH predictions are
  very close, while the TI approach shows noticeable deviations. The limited convergence of the
  TI calculations is manifested in some loss of intensity in the high-frequency part of the spectrum.
  This effect can be attributed to the truncation of the integral calculation, thereby
  neglecting a large number of very small integrals that are contributing to
  the high frequency part of the spectrum.} For Coumarin 6, \SI{94}{\percent} of the total spectral
intensity are recovered with the chosen number of computed FC integrals, {\color{black} for the spectrum without
vibrational pre-excitation.} The analytical first
moment of the spectrum is \SI{25156}{\per\centi\metre} whereas the resulting TI spectrum has a
first moment of \SI{24985}{\per\centi\metre}. {\color{black} These results may be improved using higher
  $N_{\rm max}$ thresholds (see Sec.\ II.A), at the cost of a significant increase in computational time. It is noteworthy that
  the convergence of the TI calculations is more challenging in the case of vibrational pre-excitation; this is in line with our
  previous experience for finite-temperature spectra. For further applications, it is therefore recommended to
  use the TD approach for a fast and complete calculation of the lineshape, and the TI approach
  to identify and assign the dominant bands.}

The {\color{black} spectra obtained by wavepacket propagation, using ML-MCTDH,
feature} a somewhat higher total intensity compared to the other methods and {\color{black} are} also slightly
blue-shifted with a first moment of \SI{25392}{\per\centi\metre}. In \cref{fig:c6_0k}, this 
shift has been corrected by \SI{90}{\per\centi\metre} to show the close agreement of the spectral
shape and the detailed features.
A slight discrepancy remains, {\color{black} though,} due to the different intensity distributions.
{\color{black} We attribute the deviation from the analytical TD approach -- both in terms of
  intensity and systematic frequency shifts -- to the fact that the ML-MCTDH calculations are not
  completely converged.}

The influence of vibrational pre-excitation on the electronic absorption spectrum of Coumarin 6 is
{\color{black} analyzed in further detail} in \cref{fig:c6_exptheo}, together with experimental VIPER measurements. In the wavenumber
range above \SI{1580}{\per\centi\metre} only three bands appear, corresponding to two ring modes
(labeled \emph{RM 1} and \emph{RM 2} for the low and high frequency mode, respectively) and one
carbonyl stretch mode (labeled \emph{CO}). In general, vibrational pre-excitation from the global
ground state induces additional vibronic transitions upon electronic excitation which have a lower
energy than the 0-0-transition. In particular, transitions to the vibrational
ground state of the excited electronic state (denoted the M-0-transition) become feasible. These
additional transitions induce a red-shift of the low-energy edge of the absorption band. An
efficient VIPER excitation requires a high transition probability in this frequency range. In panel
(A) of \cref{fig:c6_exptheo} the M-0 transition after excitation of the lower-frequency ring mode
(red) is clearly visible, while the corresponding transitions resulting from excitation of the
higher-frequency ring mode and the carbonyl stretch mode are much weaker. This finding is in
agreement with the experimental results in panel (C) of \cref{fig:c6_exptheo}, showing the largest
VIPER signal upon excitation of the lower-frequency ring mode.

Interestingly, the first moment of the spectrum changes only very slightly upon pre-excitation; in
case of the low-frequency ring mode it changes from \SI{25156}{\per\centi\metre} to
\SI{25082}{\per\centi\metre}. On the other hand, the standard deviation of the spectrum increases
from \SI{1744}{\per\centi\metre} to \SI{2056}{\per\centi\metre}, i.\ e.\ the spectrum gets broader
while remaining constant in energy. The additional low-frequency transitions are countered by a
more pronounced tail in the high frequency range (cf.\ panel (A) of \cref{fig:c6_exptheo}).

For completeness, we also computed spectra at finite temperature, as documented in Appendix
\ref{sec:appC}.
These show that the spectral structures are essentially unchanged, indicating that the
VIPER excitation will give very similar results at non-zero temperature. 

\subsection{7-(Diethylamino)coumarin azide {\color{black} (DEACM-N$_3$)}}

\begin{figure}[b]
    \includegraphics[width=0.95\columnwidth]{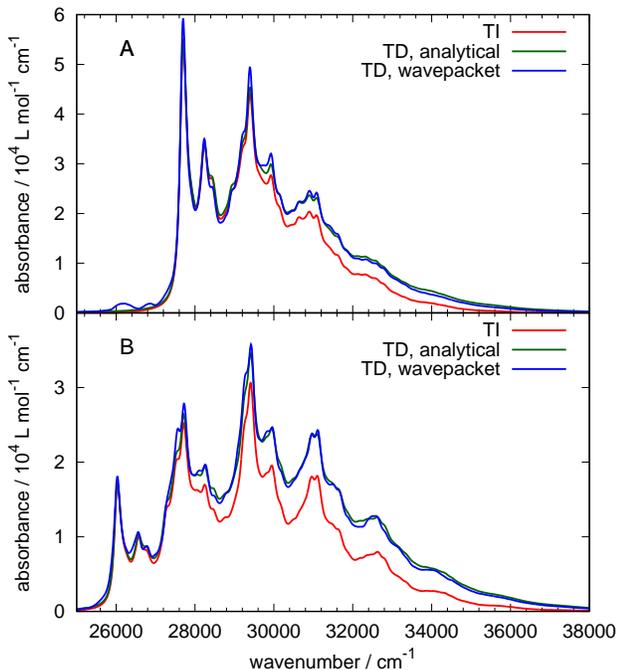}
    \caption{Computed absorption {\color{black} spectra}  of DEACM-N$_3$ in acetonitrile obtained
             from \mbox{$\omega$B97x-D}/Def2-TZVP calculations, {\color{black} (A) without vibrational pre-excitation, (B) with
                            pre-excitation of the lower-frequency ring mode (see also Fig.\ 6A).}
             The color coding is equivalent to
             \cref{fig:c6_0k}. The {\color{black} spectra} obtained from wavepacket propagation has been
             redshifted by \SI{150}{\per\centi\metre}.}\label{fig:caz_0k}
\end{figure}

\begin{figure*}
    \includegraphics[width=0.9\textwidth]{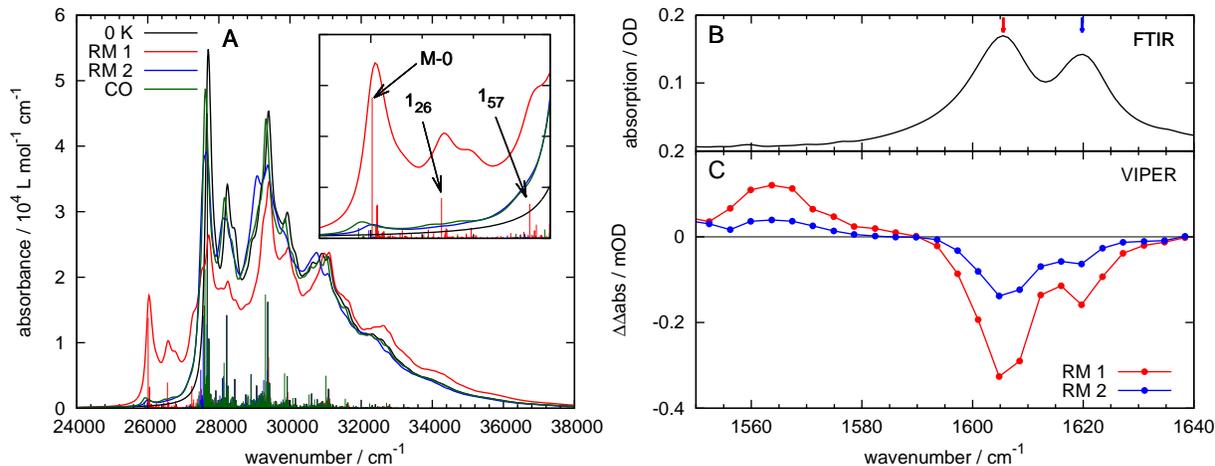}
    \caption{\textbf{(A)} Computed absorption spectra of DEACM-N$_3$ in acetonitrile using
             the analytical TD approach, {\color{black} along with stick spectra from the TI approach, based upon} \mbox{$\omega$B97x-D}/Def2-TZVP
             data. Spectra are shown without vibrational pre-excitation (black), with
             pre-excitation of the lower-frequency ring mode (red), the higher-frequency ring mode
             (blue) and the CO stretch mode (green). \textbf{(B)} Experimental FTIR absorption
             spectrum of DEACM-N$_3$ in acetonitrile. \textbf{(C)} VIPER spectra of DEACM-N$_3$ in
             acetonitrile after pre-excitation of the lower-frequency ring mode (red) and the
             higher-frequency ring mode (blue). Experimental spectra were recorded using a
             concentration of 25 mM and a layer thickness of
             250 $\mu$m.}\label{fig:caz_exptheo}
\end{figure*}

{\color{black} DEACM-N$_3$ is also a coumarin derivative, with 102 vibrational modes. Photodissociation of the azide ($N_3$) leaving group (``uncaging'') occurs on a picosecond time scale\cite{Bredenbeck2017}.} 
The $S_0$-$S_1$ transition features a pronounced $\pi$-$\pi^*$-character and is
localized to the coumarin moiety, very similar to Coumarin 6. Due to the structural similarity,
vibrational analysis of DEACM-N$_3$ yields again two distinct ring distortion modes and an intense
CO stretch mode in the relevant IR frequency range.

The computed electronic absorption spectrum of DEACM-N$_3$ is shown in \cref{fig:caz_0k},
{\color{black} again without vibrational pre-excitation (panel (A)) and with
  vibrational pre-excitation of the lower-frequency ring mode (panel (B)).} The
patterns present in the spectra of Coumarin 6 are also observed in this case. The analytical TD
method {\color{black} and the wavepacket approach are again in very good agreement and 
fully recover the high-frequency part of the spectrum,} while the convergence of the
TI method is at \SI{88}{\percent}. The first spectral moment has an analytical value of
\SI{29964}{\per\centi\metre},  {\color{black} for the spectrum without pre-excitation,}
while the TI method gives a first moment of \SI{29642}{\per\centi\metre}. {\color{black}
  As compared with Coumarin 6, the lack of convergence of the TI approach in the case of
  vibrational pre-excitation is found to be more pronounced for DEACM-N$_3$.}

The spectrum resulting from the wavepacket propagation is again
blueshifted with a first moment of \SI{30204}{\per\centi\metre} and also with a slightly higher
intensity. However, the agreement of the features of the spectral lineshape is again very good,
{\color{black} except for the spurious intensity below the 0-0 transition. The latter
  feature, together with the spectral shift and the slight deviations in intensity are most likely due to the
  incomplete convergence of the ML-MCTDH calculations, as discussed above.}

\Cref{fig:caz_exptheo} shows the computed absorption spectra of DEACM-N$_3$ under the influence of
vibrational pre-excitation {\color{black} of different modes}, together with experimental VIPER measurements. Computations show
strong additional redshifted transitions upon excitation of the lower frequency ring mode. In this
case the first moment is slightly reduced to \SI{29918}{\per\centi\metre} while the standard
deviation increases from \SI{2013}{\per\centi\metre} to \SI{2488}{\per\centi\metre}. The other
modes appear to have a much weaker effect on the spectrum. This observation is consistent with the
corresponding results from the calculations on Coumarin 6. This result again supports the
experimental findings where the measured VIPER signal has a significantly higher intensity upon
pre-excitation of the lower-frequency ring mode compared to the higher-frequency ring mode (see
panel (C) of \cref{fig:caz_exptheo}).

\subsection{\emph{para}-Hydroxyphenacyl thiocyanate {\color{black} (\emph{p}HP-SCN)}}

{\color{black} The \emph{p}HP-SCN molecule belongs to the class of para-hydroxy-phenacyl protecting groups\cite{klan_photoremovable_2012,Givens97}, with thiocyanate (SCN) as a leaving group.} 
Vibrational analysis of \emph{p}HP-SCN yields two normal modes {\color{black} (among a total of 54 modes)} in the
relevant frequency range, {\color{black} both of which are combinations of
a ring distortion mode and a carbonyl stretch mode. One of these modes is predominantly of
carbonyl stretch type, accompanied by a slight distortion of the ring,
and {\em vice versa} for the second mode. In both modes, the ring distortion and CO stretch
motions are anti-correlated in the sense that an extension (contraction) 
of the ring distortion coordinate is accompanied by a shortening (lengthening) of the C=O
bond.}
%The relative phase of these
%motions is combined with positive and negative sign, respectively, in the sense that the positive
%displacement along the ring distortion coordinate is accompanied by a shortening of the C=O bond in
%the case of the ring distortion mode and an elongation of the C=O bond in the case of the CO
%stretch mode.

The lowest singlet excited state of \emph{p}HP-SCN with non-vanishing
oscillator strength corresponds to a $\pi$-$\pi^*$-transition in the benzene
and carbonyl moiety. The excited-state structure of \emph{p}HP-SCN posed a
significant problem during the search for an excited state minimum. Depending
on the choice of density functional and solvation model, {\color{black}
  extensive} state mixing, including exchange of oscillator strength between
the state of interest and another, energetically close state, was observed,
along with strong distortions of the molecular structure.
%including dissociation of the leaving group.
{\color{black} This is an indication of nonadiabatic coupling between several states -- possibly involving
triplet states\cite{Givens97} -- }necessitating a treatment with
higher-level, {\color{black} multiconfigurational electronic
structure methods}. In the following,
results obtained with the $\omega$B97x-D density functional in THF are
reported, {\color{black} which led to a well-defined excited-state minimum structure.}

\begin{figure*}
    \includegraphics[width=0.9\textwidth]{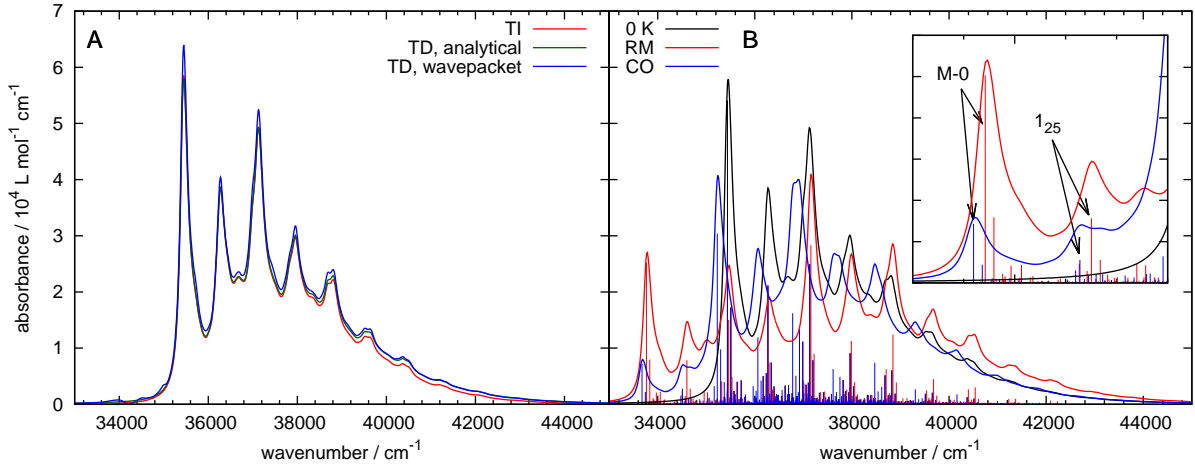}
    \caption{\textbf{(A)} Computed absorption spectra for \emph{p}HP-SCN in THF obtained from
             \mbox{$\omega$B97x-D}/Def2-TZVP calculations. The color coding is equivalent to
             \cref{fig:c6_0k}. The spectrum obtained from wavepacket propagation has been
             redshifted by \SI{45}{\per\centi\metre}. \textbf{(B)} Absorption spectra computed with
             the analytical autocorrelation function method without vibrational pre-excitation
             (black), with excitation of the ring distortion mode (red) and the CO stretch mode
             (blue). {\color{black} Stick spectra from the TI approach are also shown.}}\label{fig:php_spec}
\end{figure*}

\Cref{fig:php_spec} shows the computed spectra of \emph{p}HP-SCN. Since there are no experimental
VIPER measurements on \emph{p}HP-SCN yet, this presents an opportunity to make theoretical
predictions. In panel (A) the \SI{0}{\kelvin} spectra obtained with the different methods show
nearly quantitative agreement. The analytical first moment of the spectrum is
\SI{37582}{\per\centi\metre}, the TI and wavepacket method yield a value of
\SI{37508}{\per\centi\metre} and \SI{37726}{\per\centi\metre}, respectively. The TI method is able
to recover \SI{98}{\percent} of the total intensity.

As shown in panel (B) of \cref{fig:php_spec} vibrational pre-excitation of the ring distortion mode
again has the strongest influence on the shape of the spectrum, shifting the analytical first
moment to \SI{37601}{\per\centi\metre} while increasing the standard deviation from
\SI{1787}{\per\centi\metre} to \SI{2428}{\per\centi\metre}. However, contrary to the previous
examples, the change of the spectrum induced by excitation of the CO stretch mode is 
non-negligible. In this case, the first moment is \SI{37367}{\per\centi\metre} and the standard
deviation is increased to \SI{2080}{\per\centi\metre}. 

\section{Discussion}\label{sec:discussion}

The above results for electronic absorption spectra under the influence of vibrational
pre-excitation illustrate good agreement between the different computational approaches, and
show a pronounced mode-selectivity that is consistent with experiment. Excitation of ring
distortion modes induces substantially stronger effects on the spectrum's low-energy edge than
localized stretching modes (e.g.\ carbonyl modes). These effects can be rationalized by analyzing the
ratio of the M-0 to 0-0 transitions (cf. \cref{eq:bvec}). For the calculations detailed in
\cref{sec:results}, the dimensionless displacements $\delta$ and the intensity ratios $S$ are given
in \cref{tab:jandk}.

\begin{table}[b!]
    \centering
    \caption{Dimensionless displacements along the ground state normal modes $\delta_k$ and
             resulting transition intensity ratios $S_k$ for the relevant normal modes of the
             systems investigated here. Data for Coumarin 6 and \emph{p}HP-SCN is obtained from
             \mbox{$\omega$B97x-D}/Def2-TZVP calculations in THF while data for DEACM-N$_3$
             results from \mbox{$\omega$B97x-D}/Def2-TZVP calculations in acetonitrile.
             }\label{tab:jandk}
    \begin{tabular}{ccccccc}
        \toprule
        & \multicolumn{2}{c}{Coumarin 6} & \multicolumn{2}{c}{DEACM-N$_3$} & \multicolumn{2}{c}{\emph{p}HP-SCN} \\
        \cmidrule(lr){2-3} \cmidrule(lr){4-5} \cmidrule(lr){6-7}
        mode         & $\delta_k$ & $S_k$ & $\delta_k$ & $S_k$ & $\delta_k$ & $S_k$ \\
        \midrule
        ring mode 1  &  0.612  &  0.603  & -0.805  & -0.809  & -0.999  & -0.988  \\[1em]
        ring mode 2  & -0.179  & -0.131  & -0.237  & -0.195  &  ---    &  ---    \\[1em]
        CO stretch   & -0.053  & -0.043  &  0.214  &  0.222  & -0.554  & -0.528  \\
        \bottomrule
    \end{tabular}
\end{table}

Overall the correspondence between the values for $S_k$ and the observed spectral lineshapes is
very good. In the case of Coumarin 6, the lower-frequency ring mode has by far the highest
intensity, while the intensity of the higher-frequency ring mode is much lower, and the CO
stretch mode mode exhibits negligible intensity (cf. \cref{fig:c6_exptheo}). In the case of
DEACM-N$_3$, the higher-frequency ring mode
and the CO stretch mode have comparatively higher intensities which is also reflected in the
computed spectra (cf. \cref{fig:caz_exptheo}). In this case the CO stretch mode actually yields a
slightly higher intensity than the higher-frequency ring mode. In \emph{p}HP-SCN the difference
between the intensity ratios is not as large as in the coumarin based systems. Here the CO stretch
mode appears to be a reasonable choice as a candidate for VIPER excitation. In both coumarin based
systems the theoretical predictions agree quite well with the experimental findings regarding the
VIPER efficiency of the investigated normal
modes\cite{bredenbeck_viper2014_int,kern_steering_inprep}.

From \cref{eq:bvec} it is apparent that $\delta_k$ has a strong influence on $S_k$. In
\cref{tab:jandk}, the
values for $S$ are indeed very similar to the corresponding $\delta$-values. Additionally, in most cases
$S_k$ is lower in magnitude than $\delta_k$, the only exception being the CO stretch mode of
DEACM-N$_3$. It appears that the displacement of a mode has a much greater influence on its
suitability for VIPER excitation than the frequency change and the Duschinsky rotation which are
both contributing to the matrix term in \cref{eq:bvec}.

\begin{figure}
    \includegraphics[width=0.85\columnwidth]{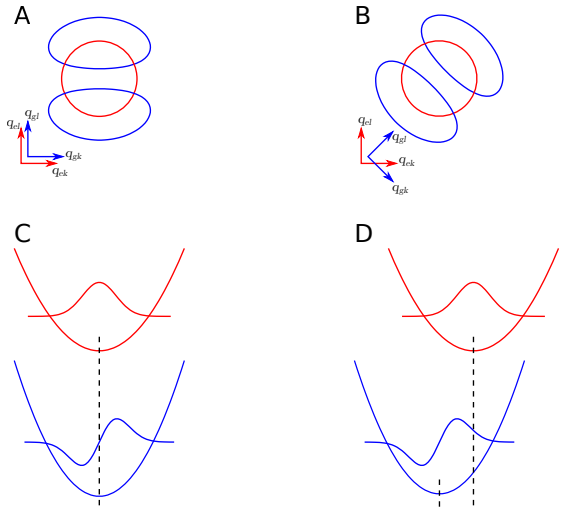}
    \caption{Schematic illustration of the influence of normal coordinates displacement vs
             Duschinsky rotation on the overlap of the pre-excited vibrational state
             {\color{black} $\ket{\bm{0}_g + 1_{gk}}$ with the vibrational ground state $\ket{\bm{0}_e}$.} \textbf{(A)}
             Contour view of the product wavefunction of {\color{black} modes $q_{gk}$ and $q_{gl}$, where $q_{gl}$ has
             been pre-excited (blue) and the corresponding wavefunction of $q_{ek}$ and $q_{el}$ (red)
             which represents the vibrational ground state in the excited electronic state. \textbf{(B)} The same but with the
             ground state modes $q_{gk}$ and $q_{gl}$} rotated w.\ r.\ t.\ their excited state
             counterparts. Obviously, the overlap is equal in cases \textbf{(A)} and \textbf{(B)}.
             \textbf{(C)} Ground and excited state PES of a pre-excited mode along with the
             corresponding wavefunctions. The minima of both potentials are coinciding.
             \textbf{(D)} The same but with a finite displacement between the PESs. Obviously, the
             overlap is significantly different in cases \textbf{(C)} and \textbf{(D)}.
             }\label{fig:rotdis}
\end{figure}

To further illustrate these points, \Cref{fig:rotdis} schematically depicts the
vibrational wavefunctions
{\color{black} $\ket{\bm{0}_g + 1_{gk}}$ and $\ket{\bm{0}_e}$} whose overlap is critical to the
VIPER effect for a simplified model system. Panel (A) shows the simplest case where the modes of
the two electronic states coincide. By symmetry the overlap integral
{\color{black} 
$\braket{\bm{0}_g + 1_{gk} | \bm{0}_e}$} must vanish, even if the frequencies of the excited state modes
differ from those of the ground state modes. Therefore, no VIPER transition can occur. Panel (B)
shows a corresponding system where the ground state modes are simply rotated with respect to the
excited state modes without any displacement. Again, by symmetry the integral vanishes. This can be
shown formally by rewriting \cref{eq:bvec} using $\bm{J} = \bm{1}$
{\color{black} 
\begin{equation}
    S_k = 2 \delta_{k}
    \left( 1 - \frac{\Omega_{gk}}{\Omega_{gk} + \Omega_{ek}} \right)
\end{equation}
}
showing that if {\color{black} $\delta_{k}$}
vanishes, so does $S_k$. Panel (C) shows a 1D cut of the system of
panel (A) along $q_l$. Clearly, if no displacement is present, the overlap integral vanishes. In
panel (D) a finite displacement between the PESs of the two electronic states is present, leading
to a nonzero overlap between the vibrational states, as expressed by \cref{eq:bvec}.

Finally, one should mention that the agreement between the different computational
methods is good, but not perfect. Importantly, the spectral features match very well. 
However, the spectra obtained by wavepacket
propagation are consistently blue-shifted relative to their respective analytical counterparts.
This effect is likely to result from numerical inaccuracies in energetic offsets, or possibly from
the limited convergence of the quantum dynamical propagation due to the finite basis size.
In addition, the spectra obtained by wavepacket propagation tend to exhibit spurious structures
below the 0-0 transition, which we also attribute to insufficient numerical convergence.  

\section{Conclusion and outlook}\label{sec:conclusion}

In this work we have presented complementary time-independent and
time-dependent approaches for the computation of vibrationally
resolved electronic absorption spectra including the effect of pre-excitation of a specific
vibrational normal mode. Within the harmonic approximation to the potential energy surfaces,
the three methods that were employed -- the time-independent \emph{FCclasses} approach and the
time-dependent analytical and numerical wavepacket approaches --    
formally coincide, and in practice yield good numerical agreement.
At this level of treatment, analytical approaches are indeed attractive and efficient
alternatives to the more general numerical wavepacket approach (see also Refs.\
[\onlinecite{huh_application_2011,huh_coherent_2012}] for related analytical developments).

In the present context, these methods were applied to three chromophores to investigate their
suitability for excitation with the VIPER mixed IR/VIS pulse sequence. We have shown that
vibrational pre-excitation leads to an intense M-0 vibronic transition if the equilibrium structure
of the excited state features a large displacement along the selected mode with respect to the
ground state equilibrium configuration. In this context the role of Duschinsky mixing appears to be
less important; in particular a purely rotated vibrational structure without any displacement is 
shown not to yield any M-0 transitions. The effect of pre-excitation is most pronounced if the
electronic transition involves the same structural region of the molecule as the pre-excited normal
mode, resulting in a strong vibronic coupling.

Follow-up work will tend to favor the time-dependent approach, in view of including effects of
Intramolecular Vibrational Redistribution (IVR) and simulating the full VIPER experiment. While the
evaluation of FC integrals and of the analytical autocorrelation function require harmonic
PESs, high-dimensional wavepacket propagation is a flexible strategy and can 
be performed on coupled surfaces line in the case of Linear (LVC) or Quadratic (QVC) Vibronic
Coupling models\cite{cederbaum_lvc_2004}, and on anharmonic surfaces as well. This opens the road
for investigations of both excitations to states exhibiting strong nonadiabatic couplings and the
vibrationally hot system's dynamics on the anharmonic ground state potential energy surface during
the time between the IR and the VIS pulse.
%Experimental observations\cite{bredenbeck_viper2014_int}
%suggest a decay of the vibrational excitation within about \SI{5}{\pico\second}.

\section*{Acknowledgement}

We thank Robert Binder for preliminary work on the Coumarin 6 system. We gratefully acknowledge
funding by the Deutsche Forschungsgemeinschaft via RTG 1986 ``Complex Scenarios of Light Control''.
{\color{black} J. C. acknowledges a fellowship provided by ``Fundaci\'on S\'eneca -- Agencia de Ciencia y Tecnolog\'ia de la Regi\'on de Murcia'' through the ``Saavedra-Fajardo''
recruitment program.}

\vspace*{1.0cm}

\appendix

\section{Derivation of the analytical autocorrelation function}\label{sec:appA}

Starting from \cref{eq:Gauss_integral_intermediate} of Sec.\ II.B.1, we introduce an orthogonal
transformation to sum and difference coordinates, $\bm{Z} = (\bm{Q}_e + \bar{\bm{Q}}_e) / \sqrt{2},
\bm{U} = (\bm{Q}_e - \bar{\bm{Q}}_e) / \sqrt{2}$, to simplify the integrals, in line
with Ref.\ [\onlinecite{baiardi_general_2013}]. This results in the following expression for the
autocorrelation function, 
\begin{widetext}
\begin{eqnarray}
    \chi_k (t) = 2 \mu_0^2 \Gamma_{gk} e^{\mathrm{i}(E_0 / \hbar + \omega_{gk})t}
    &&\sqrt{\frac{\det(\bm{\Gamma}_g) \det(\bm{a}_e)}{\pi^N (2 \pi \mathrm{i} \hbar)^N}}
    \exp \left[-\bm{K}^T \bm{\Gamma}_g \bm{K} \right] \int d\bm{Z} \int d\bm{U} \nonumber \\
    &&\times \left( K_k^2 + \sqrt{2} K_k \sum_l J_{kl} Z_l
    + \frac{1}{2} \sum_l \sum_m J_{kl} J_{km} (Z_l + U_l)(Z_m - U_m) \right) \nonumber \\
    &&\times \exp \left[ - \frac{1}{2} \bm{Z}^T \bm{D} \bm{Z}
    - \sqrt{2} \bm{\lambda}^T \bm{Z} \right]
    \exp \left[ - \frac{1}{2} \bm{U}^T \bm{C} \bm{U} \right]
\end{eqnarray}
where we used $\bm{\lambda}^T := \bm{K}^T \bm{\Gamma}_g \bm{J}$ in addition to the definitions of Eq.\ (\ref{eq:CD}).

The terms that are linear in $U_l$ and $U_m$ give no contribution to the integral and can therefore
be eliminated. Additionally, the notation can be simplified by defining the vector
$\bm{\alpha}^{(k)}$ with $\alpha^{(k)}_l = J_{kl}$ and the matrix $\bm{\beta}^{(k)}$ with
$\beta^{(k)}_{lm} = J_{kl} J_{km}$:
\begin{eqnarray}
    \chi_k (t) = 2 \mu_0^2 \Gamma_{gk} e^{\mathrm{i}(E_0 / \hbar + \omega_{gk})t}
    &&\sqrt{\frac{\det(\bm{\Gamma}_g) \det(\bm{a}_f)}{\pi^N (2 \pi \mathrm{i} \hbar)^N}}
    \exp \left[-\bm{K}^T \bm{\Gamma}_g \bm{K} \right] \int d\bm{Z} \int d\bm{U} \nonumber \\
    &&\times \left( K_k^2 + \sqrt{2} K_k \left( \bm{\alpha}^{(k)} \right)^T \bm{Z}
    + \frac{1}{2} \bm{Z}^T \bm{\beta}^{(k)} \bm{Z}
    - \frac{1}{2} \bm{U}^T \bm{\beta}^{(k)} \bm{U} \right) \nonumber \\
    &&\times \exp \left[ - \frac{1}{2} \bm{Z}^T \bm{D} \bm{Z} - \sqrt{2} \bm{\lambda}^T \bm{Z} \right]
    \exp \left[ - \frac{1}{2} \bm{U}^T \bm{C} \bm{U} \right]
\end{eqnarray}
Before proceeding, we simplify the prefactor by using
\begin{equation}
    e^{\mathrm{i} \frac{E_0}{\hbar} t} = \prod_k e^{\mathrm{i} \frac{\hbar}{2} \Gamma_{gk} t}
    = \det \left( e^{\mathrm{i} \frac{\hbar}{2} \bm{\Gamma}_g t} \right)
    = \sqrt{\frac{1}{\det \left( e^{-\mathrm{i} \hbar \bm{\Gamma}_g t} \right)}}
    \qquad \text{and} \qquad
    \bm{a}'_g (t) = \frac{2 \mathrm{i} \hbar \bm{\Gamma}_g}{e^{-\mathrm{i} \hbar \bm{\Gamma}_g t}}
\end{equation}
and split the autocorrelation function into three components:
\begin{equation}
    \chi_k (t) = 2 \Gamma_{gk} e^{\mathrm{i} \hbar \Gamma_{gk} t}
    \left[ K_k^2 \chi^0_{FC}(t) + K_k \chi^0_{k,a}(t) + \chi^0_{k,b}(t) \right]
\end{equation}
with
\begin{equation}\label{eq:ap_corr_fc0k}
    \chi^0_{FC}(t) = \mu_0^2 \sqrt{\frac{\det(\bm{a}'_g(t)\bm{a}_e(t))}
    {(2 \pi \mathrm{i} \hbar)^{2N}}}
    \exp \left[-\bm{K}^T \bm{\Gamma}_g \bm{K} \right] \int d\bm{Z} \int d\bm{U}
    \exp \left[ - \frac{1}{2} \bm{Z}^T \bm{D} \bm{Z} - \sqrt{2} \bm{\lambda}^T \bm{Z} \right]
    \exp \left[ - \frac{1}{2} \bm{U}^T \bm{C} \bm{U} \right]
\end{equation}
\begin{eqnarray}\label{eq:ap_corr_fcht0k}
    \chi^0_{k,a}(t) = \sqrt{2} \mu_0^2
    \sqrt{\frac{\det(\bm{a}'_g(t)\bm{a}_e(t))}{(2 \pi \mathrm{i} \hbar)^{2N}}}
    \exp \left[-\bm{K}^T \bm{\Gamma}_g \bm{K} \right] \int d\bm{Z} \int &&d\bm{U}
    \left( \bm{\alpha}^{(k)} \right)^T \bm{Z} \nonumber \\
    &&\exp \left[ - \frac{1}{2} \bm{Z}^T \bm{D} \bm{Z} - \sqrt{2} \bm{\lambda}^T \bm{Z} \right]
    \exp\left[ - \frac{1}{2} \bm{U}^T \bm{C} \bm{U} \right]
\end{eqnarray}
\begin{eqnarray}\label{eq:ap_corr_ht0k}
    \chi^0_{k,b}(t) = \frac{1}{2} \mu_0^2
    \sqrt{\frac{\det(\bm{a}'_g(t)\bm{a}_e(t))}{(2 \pi \mathrm{i} \hbar)^{2N}}}
    \exp \left[-\bm{K}^T \bm{\Gamma}_g \bm{K} \right] \int d\bm{Z} \int &&d\bm{U}
    \left( \bm{Z}^T \bm{\beta}^{(k)} \bm{Z} - \bm{U}^T \bm{\beta}^{(k)} \bm{U} \right) \nonumber \\
    &&\exp \left[ - \frac{1}{2} \bm{Z}^T \bm{D} \bm{Z} - \sqrt{2} \bm{\lambda}^T \bm{Z} \right]
    \exp \left[ - \frac{1}{2} \bm{U}^T \bm{C} \bm{U} \right]
\end{eqnarray}
where $\chi^0_{FC}(t)$ is the FC correlation function at \SI{0}{\kelvin} while $\chi^0_{k,a}(t)$
and $\chi^0_{k,b}(t)$ are analogous to mixed Franck-Condon/Herzberg-Teller and Herzberg-Teller terms,
respectively\cite{baiardi_general_2013}. The integrals in \cref{eq:ap_corr_fc0k} can be solved by
using the substitutions $\bm{Z}_1 = \bm{D}^{1/2} \bm{Z} + \sqrt{2} \bm{D}^{1/2} \bm{\lambda}$ and
$\bm{U}_1 = \bm{C}^{1/2} \bm{U}$, resulting in
\begin{equation}
    \chi^0_{FC}(t) = \mu_0^2 \sqrt{\frac{\det(\bm{a}'_g(t)\bm{a}_e(t))}
    {(2 \pi \mathrm{i} \hbar)^{2N} \det(\bm{CD})}}
    \exp \left[-\bm{K}^T \bm{\Gamma}_g \bm{K} \right]
    \exp \left[\bm{\lambda}^T \bm{D}^{-1} \bm{\lambda} \right]
    \int d\bm{Z}_1 \exp \left[ - \frac{1}{2} \bm{Z}_1^T \bm{Z}_1 \right]
    \int d\bm{U}_1 \exp \left[ - \frac{1}{2} \bm{U}_1^T \bm{U}_1 \right]
\end{equation}
where the Gaussian integrals are known:
\begin{equation}
    \int d\bm{Z}_1 \exp \left[ - \frac{1}{2} \bm{Z}_1^T \bm{Z}_1 \right]
    = \int d\bm{U}_1 \exp \left[ - \frac{1}{2} \bm{U}_1^T \bm{U}_1 \right]
    = (2 \pi)^{N/2}
\end{equation}
%\end{widetext}
So we finally arrive at
\begin{equation}
    \chi^0_{FC}(t) = \mu_0^2 \sqrt{\frac{\det(\bm{a}'_g(t)\bm{a}_e(t))}
    {(\mathrm{i} \hbar)^{2N} \det(\bm{CD})}}
    \exp \left[-\bm{K}^T \bm{\Gamma}_g \bm{K} + \bm{\lambda}^T \bm{D}^{-1} \bm{\lambda} \right].
\end{equation}
%\begin{eqnarray}
%    \chi^0_{FC}(t) = \mu_0^2 &&\sqrt{\frac{\det(\bm{a}'_g(t)\bm{a}_e(t))}
%    {(\mathrm{i} \hbar)^{2N} \det(\bm{CD})}} \nonumber \\
%    &&\exp \left[-\bm{K}^T \bm{\Gamma}_g \bm{K} + \bm{\lambda}^T \bm{D}^{-1} \bm{\lambda} \right].
%\end{eqnarray}
\end{widetext}
The same technique can be applied to the integrals in Eqs.\ \ref{eq:ap_corr_fcht0k} and
\ref{eq:ap_corr_ht0k} to obtain
\begin{equation}
    \chi^0_{k,a}(t) = -2 \chi^0_{FC}(t) \left( \bm{\alpha}^{(k)} \right)^T \bm{D}^{-1}
    \bm{\lambda}
\end{equation}
and
\begin{eqnarray}
    \chi^0_{k,b}(t) = \frac{1}{2} \chi^0_{FC}(t) &&\left(
    2 \bm{\lambda}^T \bm{D}^{-1} \bm{\beta}^{(k)} \bm{D}^{-1} \bm{\lambda}\right. \nonumber \\
    &&\left. + \mathrm{Tr}[\bm{\beta}^{(k)} (\bm{D}^{-1} - \bm{C}^{-1})] \right)
\end{eqnarray}
%\end{widetext}
The above expression is equivalent to Eq.\ (\ref{eq:chik_final}) of the main text, when taking into
account the additional definitions specified in \cref{sec:auto}. 

\section{Construction of the ML-MCTDH tree}\label{sec:appB}

\begin{figure}[h]
    \includegraphics[width=0.8\columnwidth]{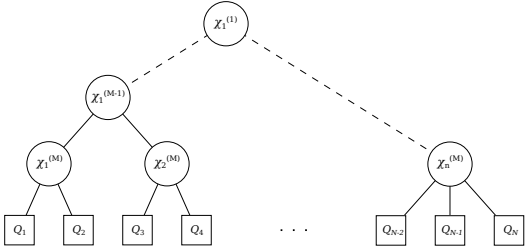}
    \caption{Schematic representation of the multilayer tree structure where $N$ normal modes are
             arranged into $M$ layers.}\label{fig:mltree}
\end{figure}

Within the ML-MCTDH framework, the wavefunction is defined in terms of a tree
structure consisting of combinations of single-particle functions (SPFs). In \cref{fig:mltree},
the structure of the multi-layer tree is illustrated. The tree was
constructed by first ordering the normal modes with increasing frequency.
{\color{black} Typically, $M=6$ layers were employed, with the number of SPFs varying from
  $n_{\rm SPF}$=3 to $n_{\rm SPF}$=8 SPFs per subspace in each layer. In the last layer, the single-particle
  functions are represented by a harmonic-oscillator (HO) Discrete-Variable Representation (DVR) with 10-20 DVR points.}  
  
  The lowest-layer particles were built by combining adjacent pairs of normal modes
into one particle, creating two-dimensional subspaces. If the total number of
normal modes is odd, the last particle contains three modes instead of two.
The lowest-level particles are then again combined pairwise, forming the
particles of the next higher layer. This process is continued until only two
or three particles remain, which then constitute the top layer.

\section{Spectra at finite temperature}\label{sec:appC}

\begin{figure}[h]
    \includegraphics[width=0.9\columnwidth]{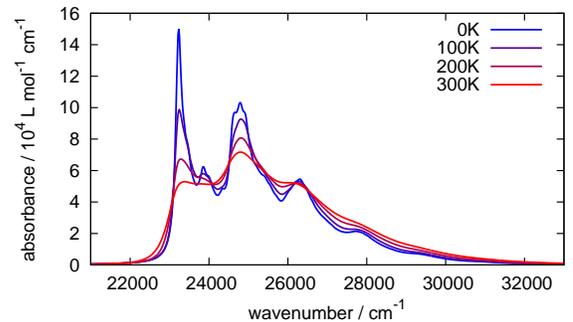}
    \caption{Computed spectra of Coumarin 6 in THF at different temperatures obtained from
             \mbox{$\omega$B97x-D}/Def2-TZVP calculations using the analytic autocorrelation
             function approach.}\label{fig:c6thermal}
\end{figure}

{\color{black} Our treatment of the VIPER excitation} assumes a \SI{0}{\kelvin}
initial state because the frequencies of the vibrational modes that are involved in the vibrational
pre-excitation are much higher than $kT$ at room temperature. Nonetheless, spectra at finite
temperature were computed, {\color{black} using the analytic TD approach\cite{tdspec_book,avila_ferrer_first-principle_2014,baiardi_general_2013,Pollak2004,Pollak2008,Lin2003,YanMukamel86,Mukamel85,Kubo55,huh_coherent_2012,Barone2014}, } to check whether these spectra are simply broadened with respect to the
zero-temperature case, or whether any additional vibronic transitions from excited vibrational
states arise. \Cref{fig:c6thermal} shows computed absorption spectra of Coumarin 6 confirming that
increasing the temperature simply leads to a broadening of the vibronic fine-structure and no
additional lower-energy transitions are observed.

\begin{thebibliography}{42}%
\makeatletter
\providecommand \@ifxundefined [1]{%
 \@ifx{#1\undefined}
}%
\providecommand \@ifnum [1]{%
 \ifnum #1\expandafter \@firstoftwo
 \else \expandafter \@secondoftwo
 \fi
}%
\providecommand \@ifx [1]{%
 \ifx #1\expandafter \@firstoftwo
 \else \expandafter \@secondoftwo
 \fi
}%
\providecommand \natexlab [1]{#1}%
\providecommand \enquote  [1]{``#1''}%
\providecommand \bibnamefont  [1]{#1}%
\providecommand \bibfnamefont [1]{#1}%
\providecommand \citenamefont [1]{#1}%
\providecommand \href@noop [0]{\@secondoftwo}%
\providecommand \href [0]{\begingroup \@sanitize@url \@href}%
\providecommand \@href[1]{\@@startlink{#1}\@@href}%
\providecommand \@@href[1]{\endgroup#1\@@endlink}%
\providecommand \@sanitize@url [0]{\catcode `\\12\catcode `\$12\catcode
  `\&12\catcode `\#12\catcode `\^12\catcode `\_12\catcode `\%12\relax}%
\providecommand \@@startlink[1]{}%
\providecommand \@@endlink[0]{}%
\providecommand \url  [0]{\begingroup\@sanitize@url \@url }%
\providecommand \@url [1]{\endgroup\@href {#1}{\urlprefix }}%
\providecommand \urlprefix  [0]{URL }%
\providecommand \Eprint [0]{\href }%
\providecommand \doibase [0]{http://dx.doi.org/}%
\providecommand \selectlanguage [0]{\@gobble}%
\providecommand \bibinfo  [0]{\@secondoftwo}%
\providecommand \bibfield  [0]{\@secondoftwo}%
\providecommand \translation [1]{[#1]}%
\providecommand \BibitemOpen [0]{}%
\providecommand \bibitemStop [0]{}%
\providecommand \bibitemNoStop [0]{.\EOS\space}%
\providecommand \EOS [0]{\spacefactor3000\relax}%
\providecommand \BibitemShut  [1]{\csname bibitem#1\endcsname}%
\let\auto@bib@innerbib\@empty
%</preamble>
\bibitem [{\citenamefont {van Wilderen}\ and\ \citenamefont
  {Bredenbeck}(2015)}]{vanwilderen_ultrafast_2015}%
  \BibitemOpen
  \bibfield  {author} {\bibinfo {author} {\bibfnamefont {L.~J. G.~W.}\
  \bibnamefont {van Wilderen}}\ and\ \bibinfo {author} {\bibfnamefont
  {J.}~\bibnamefont {Bredenbeck}},\ }\href@noop {} {\bibfield  {journal}
  {\bibinfo  {journal} {Angew. Chem. Int. Ed.}\ }\textbf {\bibinfo {volume}
  {54}},\ \bibinfo {pages} {11624} (\bibinfo {year} {2015})}\BibitemShut
  {NoStop}%
\bibitem [{\citenamefont {van Wilderen}, \citenamefont {Messmer},\ and\
  \citenamefont {Bredenbeck}(2014)}]{bredenbeck_viper2014_int}%
  \BibitemOpen
  \bibfield  {author} {\bibinfo {author} {\bibfnamefont {L.~J. G.~W.}\
  \bibnamefont {van Wilderen}}, \bibinfo {author} {\bibfnamefont {A.~T.}\
  \bibnamefont {Messmer}}, \ and\ \bibinfo {author} {\bibfnamefont
  {J.}~\bibnamefont {Bredenbeck}},\ }\href@noop {} {\bibfield  {journal}
  {\bibinfo  {journal} {Angew. Chem. Int. Ed.}\ }\textbf {\bibinfo {volume}
  {53}},\ \bibinfo {pages} {2667} (\bibinfo {year} {2014})}\BibitemShut
  {NoStop}%
\bibitem [{\citenamefont {Kern-Michler}\ \emph {et~al.}()\citenamefont
  {Kern-Michler}, \citenamefont {Neumann}, \citenamefont {Mielke},
  \citenamefont {van Wilderen}, \citenamefont {Reinfelds}, \citenamefont {von
  Cosel}, \citenamefont {Heckel}, \citenamefont {Burghardt},\ and\
  \citenamefont {Bredenbeck}}]{kern_steering_inprep}%
  \BibitemOpen
  \bibfield  {author} {\bibinfo {author} {\bibfnamefont {D.}~\bibnamefont
  {Kern-Michler}}, \bibinfo {author} {\bibfnamefont {C.}~\bibnamefont
  {Neumann}}, \bibinfo {author} {\bibfnamefont {N.}~\bibnamefont {Mielke}},
  \bibinfo {author} {\bibfnamefont {L.~J. G.~W.}\ \bibnamefont {van Wilderen}},
  \bibinfo {author} {\bibfnamefont {M.}~\bibnamefont {Reinfelds}}, \bibinfo
  {author} {\bibfnamefont {J.}~\bibnamefont {von Cosel}}, \bibinfo {author}
  {\bibfnamefont {A.}~\bibnamefont {Heckel}}, \bibinfo {author} {\bibfnamefont
  {I.}~\bibnamefont {Burghardt}}, \ and\ \bibinfo {author} {\bibfnamefont
  {J.}~\bibnamefont {Bredenbeck}},\ }\href@noop {} {\ }\bibinfo {note} {In
  preparation}\BibitemShut {NoStop}%
\bibitem [{\citenamefont {Klán}\ \emph {et~al.}(2012)\citenamefont {Klán},
  \citenamefont {Šolomek}, \citenamefont {Bochet}, \citenamefont {Blanc},
  \citenamefont {Givens}, \citenamefont {Rubina}, \citenamefont {Popik},
  \citenamefont {Kostikov},\ and\ \citenamefont
  {Wirz}}]{klan_photoremovable_2012}%
  \BibitemOpen
  \bibfield  {author} {\bibinfo {author} {\bibfnamefont {P.}~\bibnamefont
  {Klán}}, \bibinfo {author} {\bibfnamefont {T.}~\bibnamefont {Šolomek}},
  \bibinfo {author} {\bibfnamefont {C.~G.}\ \bibnamefont {Bochet}}, \bibinfo
  {author} {\bibfnamefont {A.}~\bibnamefont {Blanc}}, \bibinfo {author}
  {\bibfnamefont {R.}~\bibnamefont {Givens}}, \bibinfo {author} {\bibfnamefont
  {M.}~\bibnamefont {Rubina}}, \bibinfo {author} {\bibfnamefont
  {V.}~\bibnamefont {Popik}}, \bibinfo {author} {\bibfnamefont
  {A.}~\bibnamefont {Kostikov}}, \ and\ \bibinfo {author} {\bibfnamefont
  {J.}~\bibnamefont {Wirz}},\ }\href
  {http://pubs.acs.org/doi/abs/10.1021/cr300177k} {\bibfield  {journal}
  {\bibinfo  {journal} {Chem. Rev.}\ }\textbf {\bibinfo {volume} {113}},\
  \bibinfo {pages} {119} (\bibinfo {year} {2012})}\BibitemShut {NoStop}%
\bibitem [{\citenamefont {Park}\ and\ \citenamefont {Givens}(1997)}]{Givens97}%
  \BibitemOpen
  \bibfield  {author} {\bibinfo {author} {\bibfnamefont {C.}~\bibnamefont
  {Park}}\ and\ \bibinfo {author} {\bibfnamefont {R.}~\bibnamefont {Givens}},\
  }\href@noop {} {\bibfield  {journal} {\bibinfo  {journal} {J. Am. Chem.
  Soc.}\ }\textbf {\bibinfo {volume} {119}},\ \bibinfo {pages} {2453} (\bibinfo
  {year} {1997})}\BibitemShut {NoStop}%
\bibitem [{\citenamefont {{van Wilderen}}\ \emph {et~al.}(2017)\citenamefont
  {{van Wilderen}}, \citenamefont {Neumann}, \citenamefont
  {{Rodrigues-Correia}}, \citenamefont {{Kern-Michler}}, \citenamefont
  {Mielke}, \citenamefont {Reinfelds}, \citenamefont {Heckel},\ and\
  \citenamefont {Bredenbeck}}]{Bredenbeck2017}%
  \BibitemOpen
  \bibfield  {author} {\bibinfo {author} {\bibfnamefont {L.~J. G.~W.}\
  \bibnamefont {{van Wilderen}}}, \bibinfo {author} {\bibfnamefont
  {C.}~\bibnamefont {Neumann}}, \bibinfo {author} {\bibfnamefont
  {A.}~\bibnamefont {{Rodrigues-Correia}}}, \bibinfo {author} {\bibfnamefont
  {D.}~\bibnamefont {{Kern-Michler}}}, \bibinfo {author} {\bibfnamefont
  {N.}~\bibnamefont {Mielke}}, \bibinfo {author} {\bibfnamefont
  {M.}~\bibnamefont {Reinfelds}}, \bibinfo {author} {\bibfnamefont
  {A.}~\bibnamefont {Heckel}}, \ and\ \bibinfo {author} {\bibfnamefont
  {J.}~\bibnamefont {Bredenbeck}},\ }\href@noop {} {\bibfield  {journal}
  {\bibinfo  {journal} {Phys. Chem. Chem. Phys.}\ }\textbf {\bibinfo {volume}
  {19}},\ \bibinfo {pages} {6487} (\bibinfo {year} {2017})}\BibitemShut
  {NoStop}%
\bibitem [{\citenamefont {Santoro}\ \emph
  {et~al.}(2007{\natexlab{a}})\citenamefont {Santoro}, \citenamefont {Improta},
  \citenamefont {Lami}, \citenamefont {Bloino},\ and\ \citenamefont
  {Barone}}]{santoro_effective_2007}%
  \BibitemOpen
  \bibfield  {author} {\bibinfo {author} {\bibfnamefont {F.}~\bibnamefont
  {Santoro}}, \bibinfo {author} {\bibfnamefont {R.}~\bibnamefont {Improta}},
  \bibinfo {author} {\bibfnamefont {A.}~\bibnamefont {Lami}}, \bibinfo {author}
  {\bibfnamefont {J.}~\bibnamefont {Bloino}}, \ and\ \bibinfo {author}
  {\bibfnamefont {V.}~\bibnamefont {Barone}},\ }\href
  {http://scitation.aip.org/content/aip/journal/jcp/126/8/10.1063/1.2437197}
  {\bibfield  {journal} {\bibinfo  {journal} {J. Chem. Phys.}\ }\textbf
  {\bibinfo {volume} {126}},\ \bibinfo {pages} {084509} (\bibinfo {year}
  {2007}{\natexlab{a}})}\BibitemShut {NoStop}%
\bibitem [{\citenamefont {Santoro}\ \emph
  {et~al.}(2007{\natexlab{b}})\citenamefont {Santoro}, \citenamefont {Lami},
  \citenamefont {Improta},\ and\ \citenamefont
  {Barone}}]{santoro_effective_2007-1}%
  \BibitemOpen
  \bibfield  {author} {\bibinfo {author} {\bibfnamefont {F.}~\bibnamefont
  {Santoro}}, \bibinfo {author} {\bibfnamefont {A.}~\bibnamefont {Lami}},
  \bibinfo {author} {\bibfnamefont {R.}~\bibnamefont {Improta}}, \ and\
  \bibinfo {author} {\bibfnamefont {V.}~\bibnamefont {Barone}},\ }\href
  {http://scitation.aip.org/content/aip/journal/jcp/126/18/10.1063/1.2721539}
  {\bibfield  {journal} {\bibinfo  {journal} {J. Chem. Phys.}\ }\textbf
  {\bibinfo {volume} {126}},\ \bibinfo {pages} {184102} (\bibinfo {year}
  {2007}{\natexlab{b}})}\BibitemShut {NoStop}%
\bibitem [{\citenamefont {Santoro}\ \emph {et~al.}(2008)\citenamefont
  {Santoro}, \citenamefont {Lami}, \citenamefont {Improta}, \citenamefont
  {Bloino},\ and\ \citenamefont {Barone}}]{santoro_effective_2008}%
  \BibitemOpen
  \bibfield  {author} {\bibinfo {author} {\bibfnamefont {F.}~\bibnamefont
  {Santoro}}, \bibinfo {author} {\bibfnamefont {A.}~\bibnamefont {Lami}},
  \bibinfo {author} {\bibfnamefont {R.}~\bibnamefont {Improta}}, \bibinfo
  {author} {\bibfnamefont {J.}~\bibnamefont {Bloino}}, \ and\ \bibinfo {author}
  {\bibfnamefont {V.}~\bibnamefont {Barone}},\ }\href
  {http://scitation.aip.org/content/aip/journal/jcp/128/22/10.1063/1.2929846}
  {\bibfield  {journal} {\bibinfo  {journal} {J. Chem. Phys.}\ }\textbf
  {\bibinfo {volume} {128}},\ \bibinfo {pages} {224311} (\bibinfo {year}
  {2008})}\BibitemShut {NoStop}%
\bibitem [{\citenamefont {Wang}\ and\ \citenamefont
  {Thoss}(2003)}]{wang_multilayer_2003}%
  \BibitemOpen
  \bibfield  {author} {\bibinfo {author} {\bibfnamefont {H.}~\bibnamefont
  {Wang}}\ and\ \bibinfo {author} {\bibfnamefont {M.}~\bibnamefont {Thoss}},\
  }\href@noop {} {\bibfield  {journal} {\bibinfo  {journal} {J. Chem. Phys.}\
  }\textbf {\bibinfo {volume} {119}},\ \bibinfo {pages} {1289} (\bibinfo {year}
  {2003})}\BibitemShut {NoStop}%
\bibitem [{\citenamefont {Vendrell}\ and\ \citenamefont
  {Meyer}(2011)}]{vendrell_multilayer_2011}%
  \BibitemOpen
  \bibfield  {author} {\bibinfo {author} {\bibfnamefont {O.}~\bibnamefont
  {Vendrell}}\ and\ \bibinfo {author} {\bibfnamefont {H.-D.}\ \bibnamefont
  {Meyer}},\ }\href {\doibase 10.1063/1.3535541} {\bibfield  {journal}
  {\bibinfo  {journal} {J. Chem. Phys.}\ }\textbf {\bibinfo {volume} {134}},\
  \bibinfo {pages} {044135} (\bibinfo {year} {2011})}\BibitemShut {NoStop}%
\bibitem [{\citenamefont {Meyer}, \citenamefont {Manthe},\ and\ \citenamefont
  {Cederbaum}(1990)}]{meyer_multi-configurational_1990}%
  \BibitemOpen
  \bibfield  {author} {\bibinfo {author} {\bibfnamefont {H.-D.}\ \bibnamefont
  {Meyer}}, \bibinfo {author} {\bibfnamefont {U.}~\bibnamefont {Manthe}}, \
  and\ \bibinfo {author} {\bibfnamefont {L.~S.}\ \bibnamefont {Cederbaum}},\
  }\href@noop {} {\bibfield  {journal} {\bibinfo  {journal} {Chem. Phys.
  Lett.}\ }\textbf {\bibinfo {volume} {165}},\ \bibinfo {pages} {73} (\bibinfo
  {year} {1990})}\BibitemShut {NoStop}%
\bibitem [{\citenamefont {Beck}\ \emph {et~al.}(2000)\citenamefont {Beck},
  \citenamefont {Jäckle}, \citenamefont {Worth},\ and\ \citenamefont
  {Meyer}}]{beck_multiconfiguration_2000}%
  \BibitemOpen
  \bibfield  {author} {\bibinfo {author} {\bibfnamefont {M.~H.}\ \bibnamefont
  {Beck}}, \bibinfo {author} {\bibfnamefont {A.}~\bibnamefont {Jäckle}},
  \bibinfo {author} {\bibfnamefont {G.~A.}\ \bibnamefont {Worth}}, \ and\
  \bibinfo {author} {\bibfnamefont {H.-D.}\ \bibnamefont {Meyer}},\ }\href@noop
  {} {\bibfield  {journal} {\bibinfo  {journal} {Phys. Rep.}\ }\textbf
  {\bibinfo {volume} {324}},\ \bibinfo {pages} {1} (\bibinfo {year}
  {2000})}\BibitemShut {NoStop}%
\bibitem [{\citenamefont {Worth}\ \emph {et~al.}(2008)\citenamefont {Worth},
  \citenamefont {Meyer}, \citenamefont {Köppel}, \citenamefont {Cederbaum},\
  and\ \citenamefont {Burghardt}}]{worth_using_2008}%
  \BibitemOpen
  \bibfield  {author} {\bibinfo {author} {\bibfnamefont {G.~A.}\ \bibnamefont
  {Worth}}, \bibinfo {author} {\bibfnamefont {H.-D.}\ \bibnamefont {Meyer}},
  \bibinfo {author} {\bibfnamefont {H.}~\bibnamefont {Köppel}}, \bibinfo
  {author} {\bibfnamefont {L.~S.}\ \bibnamefont {Cederbaum}}, \ and\ \bibinfo
  {author} {\bibfnamefont {I.}~\bibnamefont {Burghardt}},\ }\href@noop {}
  {\bibfield  {journal} {\bibinfo  {journal} {Int. Rev. Phys. Chem.}\ }\textbf
  {\bibinfo {volume} {27}},\ \bibinfo {pages} {569} (\bibinfo {year}
  {2008})}\BibitemShut {NoStop}%
\bibitem [{\citenamefont {Duschinsky}(1937)}]{duschinsky37}%
  \BibitemOpen
  \bibfield  {author} {\bibinfo {author} {\bibfnamefont {F.}~\bibnamefont
  {Duschinsky}},\ }\href@noop {} {\bibfield  {journal} {\bibinfo  {journal}
  {Acta Physicochim. U. R. S. S.}\ }\textbf {\bibinfo {volume} {7}},\ \bibinfo
  {pages} {551} (\bibinfo {year} {1937})}\BibitemShut {NoStop}%
\bibitem [{\citenamefont {Tannor}(2007)}]{tannor_tdqm}%
  \BibitemOpen
  \bibfield  {author} {\bibinfo {author} {\bibfnamefont {D.~J.}\ \bibnamefont
  {Tannor}},\ }\href@noop {} {\emph {\bibinfo {title} {Introduction to quantum
  mechanics: A time-dependent perspective}}},\ \bibinfo {edition} {1st}\ ed.\
  (\bibinfo  {publisher} {University Science Books},\ \bibinfo {address}
  {USA},\ \bibinfo {year} {2007})\BibitemShut {NoStop}%
\bibitem [{\citenamefont {Santoro}()}]{fcclassesprog}%
  \BibitemOpen
  \bibfield  {author} {\bibinfo {author} {\bibfnamefont {F.}~\bibnamefont
  {Santoro}},\ }\href {http://village.pi.iccom.cnr.it/Software} {\enquote
  {\bibinfo {title} {{FCClasses}, a {Fortran} 77 code},}\ }\bibinfo {note}
  {Available at http://village.pi.iccom.cnr.it/Software}\BibitemShut {NoStop}%
\bibitem [{\citenamefont {Santoro}\ and\ \citenamefont
  {Lami}(2012)}]{tdspec_book}%
  \BibitemOpen
  \bibfield  {author} {\bibinfo {author} {\bibfnamefont {F.}~\bibnamefont
  {Santoro}}\ and\ \bibinfo {author} {\bibfnamefont {A.}~\bibnamefont {Lami}},\
  }in\ \href@noop {} {\emph {\bibinfo {booktitle} {Computational Strategies for
  Spectroscopy: From Small Molecules to Nano Systems}}},\ \bibinfo {editor}
  {edited by\ \bibinfo {editor} {\bibfnamefont {V.}~\bibnamefont {Barone}}}\
  (\bibinfo  {publisher} {John Wiley \& Sons, Inc.},\ \bibinfo {year} {2012})\
  Chap.~\bibinfo {chapter} {10}, pp.\ \bibinfo {pages} {475--516}\BibitemShut
  {NoStop}%
\bibitem [{\citenamefont {Avila~Ferrer}\ \emph {et~al.}(2014)\citenamefont
  {Avila~Ferrer}, \citenamefont {Cerezo}, \citenamefont {Soto}, \citenamefont
  {Improta},\ and\ \citenamefont
  {Santoro}}]{avila_ferrer_first-principle_2014}%
  \BibitemOpen
  \bibfield  {author} {\bibinfo {author} {\bibfnamefont {F.~J.}\ \bibnamefont
  {Avila~Ferrer}}, \bibinfo {author} {\bibfnamefont {J.}~\bibnamefont
  {Cerezo}}, \bibinfo {author} {\bibfnamefont {J.}~\bibnamefont {Soto}},
  \bibinfo {author} {\bibfnamefont {R.}~\bibnamefont {Improta}}, \ and\
  \bibinfo {author} {\bibfnamefont {F.}~\bibnamefont {Santoro}},\ }\href
  {\doibase 10.1016/j.comptc.2014.03.003} {\bibfield  {journal} {\bibinfo
  {journal} {Comput. Theor. Chem.}\ }\textbf {\bibinfo {volume}
  {1040–1041}},\ \bibinfo {pages} {328} (\bibinfo {year} {2014})}\BibitemShut
  {NoStop}%
\bibitem [{\citenamefont {Baiardi}, \citenamefont {Bloino},\ and\ \citenamefont
  {Barone}(2013)}]{baiardi_general_2013}%
  \BibitemOpen
  \bibfield  {author} {\bibinfo {author} {\bibfnamefont {A.}~\bibnamefont
  {Baiardi}}, \bibinfo {author} {\bibfnamefont {J.}~\bibnamefont {Bloino}}, \
  and\ \bibinfo {author} {\bibfnamefont {V.}~\bibnamefont {Barone}},\ }\href
  {\doibase 10.1021/ct400450k} {\bibfield  {journal} {\bibinfo  {journal} {J.
  Chem. Theory Comput.}\ }\textbf {\bibinfo {volume} {9}},\ \bibinfo {pages}
  {4097} (\bibinfo {year} {2013})}\BibitemShut {NoStop}%
\bibitem [{\citenamefont {Ianconescu}\ and\ \citenamefont
  {Pollak}(2004)}]{Pollak2004}%
  \BibitemOpen
  \bibfield  {author} {\bibinfo {author} {\bibfnamefont {E.}~\bibnamefont
  {Ianconescu}}\ and\ \bibinfo {author} {\bibfnamefont {E.}~\bibnamefont
  {Pollak}},\ }\href@noop {} {\bibfield  {journal} {\bibinfo  {journal} {J.
  Phys. Chem. A}\ }\textbf {\bibinfo {volume} {108}},\ \bibinfo {pages} {7778}
  (\bibinfo {year} {2004})}\BibitemShut {NoStop}%
\bibitem [{\citenamefont {Tatchen}\ and\ \citenamefont
  {Pollak}(2008)}]{Pollak2008}%
  \BibitemOpen
  \bibfield  {author} {\bibinfo {author} {\bibfnamefont {J.}~\bibnamefont
  {Tatchen}}\ and\ \bibinfo {author} {\bibfnamefont {E.}~\bibnamefont
  {Pollak}},\ }\href@noop {} {\bibfield  {journal} {\bibinfo  {journal} {J.
  Chem. Phys.}\ }\textbf {\bibinfo {volume} {128}},\ \bibinfo {pages} {164303}
  (\bibinfo {year} {2008})}\BibitemShut {NoStop}%
\bibitem [{\citenamefont {Tang}, \citenamefont {Lee},\ and\ \citenamefont
  {Lin}(2003)}]{Lin2003}%
  \BibitemOpen
  \bibfield  {author} {\bibinfo {author} {\bibfnamefont {J.}~\bibnamefont
  {Tang}}, \bibinfo {author} {\bibfnamefont {M.~T.}\ \bibnamefont {Lee}}, \
  and\ \bibinfo {author} {\bibfnamefont {S.~H.}\ \bibnamefont {Lin}},\
  }\href@noop {} {\bibfield  {journal} {\bibinfo  {journal} {J. Chem. Phys.}\
  }\textbf {\bibinfo {volume} {119}},\ \bibinfo {pages} {7188} (\bibinfo {year}
  {2003})}\BibitemShut {NoStop}%
\bibitem [{\citenamefont {Yan}\ and\ \citenamefont
  {Mukamel}(1986)}]{YanMukamel86}%
  \BibitemOpen
  \bibfield  {author} {\bibinfo {author} {\bibfnamefont {Y.}~\bibnamefont
  {Yan}}\ and\ \bibinfo {author} {\bibfnamefont {S.}~\bibnamefont {Mukamel}},\
  }\href@noop {} {\bibfield  {journal} {\bibinfo  {journal} {J. Chem. Phys.}\
  }\textbf {\bibinfo {volume} {85}},\ \bibinfo {pages} {5908} (\bibinfo {year}
  {1986})}\BibitemShut {NoStop}%
\bibitem [{\citenamefont {Mukamel}, \citenamefont {Abe},\ and\ \citenamefont
  {Islampour}(1985)}]{Mukamel85}%
  \BibitemOpen
  \bibfield  {author} {\bibinfo {author} {\bibfnamefont {S.}~\bibnamefont
  {Mukamel}}, \bibinfo {author} {\bibfnamefont {S.}~\bibnamefont {Abe}}, \ and\
  \bibinfo {author} {\bibfnamefont {R.}~\bibnamefont {Islampour}},\ }\href@noop
  {} {\bibfield  {journal} {\bibinfo  {journal} {J. Phys. Chem.}\ }\textbf
  {\bibinfo {volume} {89}},\ \bibinfo {pages} {201} (\bibinfo {year}
  {1985})}\BibitemShut {NoStop}%
\bibitem [{\citenamefont {Kubo}\ and\ \citenamefont {Toyozawa}(1955)}]{Kubo55}%
  \BibitemOpen
  \bibfield  {author} {\bibinfo {author} {\bibfnamefont {R.}~\bibnamefont
  {Kubo}}\ and\ \bibinfo {author} {\bibfnamefont {Y.}~\bibnamefont
  {Toyozawa}},\ }\href@noop {} {\bibfield  {journal} {\bibinfo  {journal}
  {Prog. Theor. Phys.}\ }\textbf {\bibinfo {volume} {13}},\ \bibinfo {pages}
  {160} (\bibinfo {year} {1955})}\BibitemShut {NoStop}%
\bibitem [{\citenamefont {Huh}\ and\ \citenamefont
  {Berger}(2012)}]{huh_coherent_2012}%
  \BibitemOpen
  \bibfield  {author} {\bibinfo {author} {\bibfnamefont {J.}~\bibnamefont
  {Huh}}\ and\ \bibinfo {author} {\bibfnamefont {R.}~\bibnamefont {Berger}},\
  }\href@noop {} {\bibfield  {journal} {\bibinfo  {journal} {J. Phys. Conf.
  Ser.}\ }\textbf {\bibinfo {volume} {380}},\ \bibinfo {pages} {012019}
  (\bibinfo {year} {2012})}\BibitemShut {NoStop}%
\bibitem [{\citenamefont {Baiardi}, \citenamefont {Bloino},\ and\ \citenamefont
  {Barone}(2014)}]{Barone2014}%
  \BibitemOpen
  \bibfield  {author} {\bibinfo {author} {\bibfnamefont {A.}~\bibnamefont
  {Baiardi}}, \bibinfo {author} {\bibfnamefont {J.}~\bibnamefont {Bloino}}, \
  and\ \bibinfo {author} {\bibfnamefont {V.}~\bibnamefont {Barone}},\
  }\href@noop {} {\bibfield  {journal} {\bibinfo  {journal} {J. Chem. Phys.}\
  }\textbf {\bibinfo {volume} {141}},\ \bibinfo {pages} {114108} (\bibinfo
  {year} {2014})}\BibitemShut {NoStop}%
\bibitem [{\citenamefont {Peng}\ \emph {et~al.}(2010)\citenamefont {Peng},
  \citenamefont {Niu}, \citenamefont {Deng},\ and\ \citenamefont
  {Shuai}}]{peng_vibration_2010}%
  \BibitemOpen
  \bibfield  {author} {\bibinfo {author} {\bibfnamefont {Q.}~\bibnamefont
  {Peng}}, \bibinfo {author} {\bibfnamefont {Y.}~\bibnamefont {Niu}}, \bibinfo
  {author} {\bibfnamefont {C.}~\bibnamefont {Deng}}, \ and\ \bibinfo {author}
  {\bibfnamefont {Z.}~\bibnamefont {Shuai}},\ }\href {\doibase
  10.1016/j.chemphys.2010.03.004} {\bibfield  {journal} {\bibinfo  {journal}
  {Chem. Phys.}\ }\textbf {\bibinfo {volume} {370}},\ \bibinfo {pages} {215}
  (\bibinfo {year} {2010})}\BibitemShut {NoStop}%
\bibitem [{\citenamefont {Borrelli}, \citenamefont {Capobianco},\ and\
  \citenamefont {Peluso}(2012)}]{Peluso2012}%
  \BibitemOpen
  \bibfield  {author} {\bibinfo {author} {\bibfnamefont {R.}~\bibnamefont
  {Borrelli}}, \bibinfo {author} {\bibfnamefont {A.}~\bibnamefont
  {Capobianco}}, \ and\ \bibinfo {author} {\bibfnamefont {A.}~\bibnamefont
  {Peluso}},\ }\href@noop {} {\bibfield  {journal} {\bibinfo  {journal} {J.
  Phys. Chem. A}\ }\textbf {\bibinfo {volume} {116}},\ \bibinfo {pages} {9934}
  (\bibinfo {year} {2012})}\BibitemShut {NoStop}%
\bibitem [{\citenamefont {Biczysko}\ \emph {et~al.}(2012)\citenamefont
  {Biczysko}, \citenamefont {Bloino}, \citenamefont {Santoro},\ and\
  \citenamefont {Barone}}]{tispec_book}%
  \BibitemOpen
  \bibfield  {author} {\bibinfo {author} {\bibfnamefont {M.}~\bibnamefont
  {Biczysko}}, \bibinfo {author} {\bibfnamefont {J.}~\bibnamefont {Bloino}},
  \bibinfo {author} {\bibfnamefont {F.}~\bibnamefont {Santoro}}, \ and\
  \bibinfo {author} {\bibfnamefont {V.}~\bibnamefont {Barone}},\ }in\
  \href@noop {} {\emph {\bibinfo {booktitle} {Computational Strategies for
  Spectroscopy: From Small Molecules to Nano Systems}}},\ \bibinfo {editor}
  {edited by\ \bibinfo {editor} {\bibfnamefont {V.}~\bibnamefont {Barone}}}\
  (\bibinfo  {publisher} {John Wiley \& Sons, Inc.},\ \bibinfo {year} {2012})\
  Chap.~\bibinfo {chapter} {8}, pp.\ \bibinfo {pages} {361--443}\BibitemShut
  {NoStop}%
\bibitem [{\citenamefont {Mukamel}(1999)}]{Mukamel}%
  \BibitemOpen
  \bibfield  {author} {\bibinfo {author} {\bibfnamefont {S.}~\bibnamefont
  {Mukamel}},\ }\href@noop {} {\emph {\bibinfo {title} {Principles of Nonlinear
  Optical Spectroscopy}}},\ \bibinfo {edition} {1st}\ ed.\ (\bibinfo
  {publisher} {Oxford University Press},\ \bibinfo {address} {Oxford},\
  \bibinfo {year} {1999})\BibitemShut {NoStop}%
\bibitem [{\citenamefont {Frisch}\ \emph {et~al.}(2013)\citenamefont {Frisch},
  \citenamefont {Trucks}, \citenamefont {Schlegel}, \citenamefont {Scuseria},
  \citenamefont {Robb}, \citenamefont {Cheeseman}, \citenamefont {Scalmani},
  \citenamefont {Barone}, \citenamefont {Mennucci}, \citenamefont {Petersson},
  \citenamefont {Nakatsuji}, \citenamefont {Caricato}, \citenamefont {Li},
  \citenamefont {Hratchian}, \citenamefont {Izmaylov}, \citenamefont {Bloino},
  \citenamefont {Zheng}, \citenamefont {Sonnenberg}, \citenamefont {Hada},
  \citenamefont {Ehara}, \citenamefont {Toyota}, \citenamefont {Fukuda},
  \citenamefont {Hasegawa}, \citenamefont {Ishida}, \citenamefont {Nakajima},
  \citenamefont {Honda}, \citenamefont {Kitao}, \citenamefont {Nakai},
  \citenamefont {Vreven}, \citenamefont {Montgomery}, \citenamefont {Peralta},
  \citenamefont {Ogliaro}, \citenamefont {Bearpark}, \citenamefont {Heyd},
  \citenamefont {Brothers}, \citenamefont {Kudin}, \citenamefont {Staroverov},
  \citenamefont {Kobayashi}, \citenamefont {Normand}, \citenamefont
  {Raghavachari}, \citenamefont {Rendell}, \citenamefont {Burant},
  \citenamefont {Iyengar}, \citenamefont {Tomasi}, \citenamefont {Cossi},
  \citenamefont {Rega}, \citenamefont {Millam}, \citenamefont {Klene},
  \citenamefont {Knox}, \citenamefont {Cross}, \citenamefont {Bakken},
  \citenamefont {Adamo}, \citenamefont {Jaramillo}, \citenamefont {Gomperts},
  \citenamefont {Stratmann}, \citenamefont {Yazyev}, \citenamefont {Austin},
  \citenamefont {Cammi}, \citenamefont {Pomelli}, \citenamefont {Ochterski},
  \citenamefont {Martin}, \citenamefont {Morokuma}, \citenamefont {Zakrzewski},
  \citenamefont {Voth}, \citenamefont {Salvador}, \citenamefont {Dannenberg},
  \citenamefont {Dapprich}, \citenamefont {Daniels}, \citenamefont {Farkas},
  \citenamefont {Foresman}, \citenamefont {Ortiz}, \citenamefont {Cioslowski},\
  and\ \citenamefont {Fox}}]{g09}%
  \BibitemOpen
  \bibfield  {author} {\bibinfo {author} {\bibfnamefont {M.~J.}\ \bibnamefont
  {Frisch}}, \bibinfo {author} {\bibfnamefont {G.~W.}\ \bibnamefont {Trucks}},
  \bibinfo {author} {\bibfnamefont {H.~B.}\ \bibnamefont {Schlegel}}, \bibinfo
  {author} {\bibfnamefont {G.~E.}\ \bibnamefont {Scuseria}}, \bibinfo {author}
  {\bibfnamefont {M.~A.}\ \bibnamefont {Robb}}, \bibinfo {author}
  {\bibfnamefont {J.~R.}\ \bibnamefont {Cheeseman}}, \bibinfo {author}
  {\bibfnamefont {G.}~\bibnamefont {Scalmani}}, \bibinfo {author}
  {\bibfnamefont {V.}~\bibnamefont {Barone}}, \bibinfo {author} {\bibfnamefont
  {B.}~\bibnamefont {Mennucci}}, \bibinfo {author} {\bibfnamefont {G.~A.}\
  \bibnamefont {Petersson}}, \bibinfo {author} {\bibfnamefont {H.}~\bibnamefont
  {Nakatsuji}}, \bibinfo {author} {\bibfnamefont {M.}~\bibnamefont {Caricato}},
  \bibinfo {author} {\bibfnamefont {X.}~\bibnamefont {Li}}, \bibinfo {author}
  {\bibfnamefont {H.~P.}\ \bibnamefont {Hratchian}}, \bibinfo {author}
  {\bibfnamefont {A.~F.}\ \bibnamefont {Izmaylov}}, \bibinfo {author}
  {\bibfnamefont {J.}~\bibnamefont {Bloino}}, \bibinfo {author} {\bibfnamefont
  {G.}~\bibnamefont {Zheng}}, \bibinfo {author} {\bibfnamefont {J.~L.}\
  \bibnamefont {Sonnenberg}}, \bibinfo {author} {\bibfnamefont
  {M.}~\bibnamefont {Hada}}, \bibinfo {author} {\bibfnamefont {M.}~\bibnamefont
  {Ehara}}, \bibinfo {author} {\bibfnamefont {K.}~\bibnamefont {Toyota}},
  \bibinfo {author} {\bibfnamefont {R.}~\bibnamefont {Fukuda}}, \bibinfo
  {author} {\bibfnamefont {J.}~\bibnamefont {Hasegawa}}, \bibinfo {author}
  {\bibfnamefont {M.}~\bibnamefont {Ishida}}, \bibinfo {author} {\bibfnamefont
  {T.}~\bibnamefont {Nakajima}}, \bibinfo {author} {\bibfnamefont
  {Y.}~\bibnamefont {Honda}}, \bibinfo {author} {\bibfnamefont
  {O.}~\bibnamefont {Kitao}}, \bibinfo {author} {\bibfnamefont
  {H.}~\bibnamefont {Nakai}}, \bibinfo {author} {\bibfnamefont
  {T.}~\bibnamefont {Vreven}}, \bibinfo {author} {\bibfnamefont {J.~A.}\
  \bibnamefont {Montgomery}, \bibfnamefont {{Jr.}}}, \bibinfo {author}
  {\bibfnamefont {J.~E.}\ \bibnamefont {Peralta}}, \bibinfo {author}
  {\bibfnamefont {F.}~\bibnamefont {Ogliaro}}, \bibinfo {author} {\bibfnamefont
  {M.}~\bibnamefont {Bearpark}}, \bibinfo {author} {\bibfnamefont {J.~J.}\
  \bibnamefont {Heyd}}, \bibinfo {author} {\bibfnamefont {E.}~\bibnamefont
  {Brothers}}, \bibinfo {author} {\bibfnamefont {K.~N.}\ \bibnamefont {Kudin}},
  \bibinfo {author} {\bibfnamefont {V.~N.}\ \bibnamefont {Staroverov}},
  \bibinfo {author} {\bibfnamefont {R.}~\bibnamefont {Kobayashi}}, \bibinfo
  {author} {\bibfnamefont {J.}~\bibnamefont {Normand}}, \bibinfo {author}
  {\bibfnamefont {K.}~\bibnamefont {Raghavachari}}, \bibinfo {author}
  {\bibfnamefont {A.}~\bibnamefont {Rendell}}, \bibinfo {author} {\bibfnamefont
  {J.~C.}\ \bibnamefont {Burant}}, \bibinfo {author} {\bibfnamefont {S.~S.}\
  \bibnamefont {Iyengar}}, \bibinfo {author} {\bibfnamefont {J.}~\bibnamefont
  {Tomasi}}, \bibinfo {author} {\bibfnamefont {M.}~\bibnamefont {Cossi}},
  \bibinfo {author} {\bibfnamefont {N.}~\bibnamefont {Rega}}, \bibinfo {author}
  {\bibfnamefont {J.~M.}\ \bibnamefont {Millam}}, \bibinfo {author}
  {\bibfnamefont {M.}~\bibnamefont {Klene}}, \bibinfo {author} {\bibfnamefont
  {J.~E.}\ \bibnamefont {Knox}}, \bibinfo {author} {\bibfnamefont {J.~B.}\
  \bibnamefont {Cross}}, \bibinfo {author} {\bibfnamefont {V.}~\bibnamefont
  {Bakken}}, \bibinfo {author} {\bibfnamefont {C.}~\bibnamefont {Adamo}},
  \bibinfo {author} {\bibfnamefont {J.}~\bibnamefont {Jaramillo}}, \bibinfo
  {author} {\bibfnamefont {R.}~\bibnamefont {Gomperts}}, \bibinfo {author}
  {\bibfnamefont {R.~E.}\ \bibnamefont {Stratmann}}, \bibinfo {author}
  {\bibfnamefont {O.}~\bibnamefont {Yazyev}}, \bibinfo {author} {\bibfnamefont
  {A.~J.}\ \bibnamefont {Austin}}, \bibinfo {author} {\bibfnamefont
  {R.}~\bibnamefont {Cammi}}, \bibinfo {author} {\bibfnamefont
  {C.}~\bibnamefont {Pomelli}}, \bibinfo {author} {\bibfnamefont {J.~W.}\
  \bibnamefont {Ochterski}}, \bibinfo {author} {\bibfnamefont {R.~L.}\
  \bibnamefont {Martin}}, \bibinfo {author} {\bibfnamefont {K.}~\bibnamefont
  {Morokuma}}, \bibinfo {author} {\bibfnamefont {V.~G.}\ \bibnamefont
  {Zakrzewski}}, \bibinfo {author} {\bibfnamefont {G.~A.}\ \bibnamefont
  {Voth}}, \bibinfo {author} {\bibfnamefont {P.}~\bibnamefont {Salvador}},
  \bibinfo {author} {\bibfnamefont {J.~J.}\ \bibnamefont {Dannenberg}},
  \bibinfo {author} {\bibfnamefont {S.}~\bibnamefont {Dapprich}}, \bibinfo
  {author} {\bibfnamefont {A.~D.}\ \bibnamefont {Daniels}}, \bibinfo {author}
  {\bibfnamefont {{\"O}.}~\bibnamefont {Farkas}}, \bibinfo {author}
  {\bibfnamefont {J.~B.}\ \bibnamefont {Foresman}}, \bibinfo {author}
  {\bibfnamefont {J.~V.}\ \bibnamefont {Ortiz}}, \bibinfo {author}
  {\bibfnamefont {J.}~\bibnamefont {Cioslowski}}, \ and\ \bibinfo {author}
  {\bibfnamefont {D.~J.}\ \bibnamefont {Fox}},\ }\href@noop {} {\enquote
  {\bibinfo {title} {Gaussian 09 {R}evision {D}.01},}\ } (\bibinfo {year}
  {2013}),\ \bibinfo {note} {{G}aussian Inc. Wallingford CT}\BibitemShut
  {NoStop}%
\bibitem [{\citenamefont {Chai}\ and\ \citenamefont
  {Head-Gordon}(2008)}]{chai_long-range_2008}%
  \BibitemOpen
  \bibfield  {author} {\bibinfo {author} {\bibfnamefont {J.-D.}\ \bibnamefont
  {Chai}}\ and\ \bibinfo {author} {\bibfnamefont {M.}~\bibnamefont
  {Head-Gordon}},\ }\href {\doibase 10.1039/B810189B} {\bibfield  {journal}
  {\bibinfo  {journal} {Phys. Chem. Chem. Phys.}\ }\textbf {\bibinfo {volume}
  {10}},\ \bibinfo {pages} {6615} (\bibinfo {year} {2008})}\BibitemShut
  {NoStop}%
\bibitem [{\citenamefont {Yanai}, \citenamefont {Tew},\ and\ \citenamefont
  {Handy}(2004)}]{yanai_new_2004}%
  \BibitemOpen
  \bibfield  {author} {\bibinfo {author} {\bibfnamefont {T.}~\bibnamefont
  {Yanai}}, \bibinfo {author} {\bibfnamefont {D.~P.}\ \bibnamefont {Tew}}, \
  and\ \bibinfo {author} {\bibfnamefont {N.~C.}\ \bibnamefont {Handy}},\ }\href
  {\doibase 10.1016/j.cplett.2004.06.011} {\bibfield  {journal} {\bibinfo
  {journal} {Chem. Phys. Lett.}\ }\textbf {\bibinfo {volume} {393}},\ \bibinfo
  {pages} {51} (\bibinfo {year} {2004})}\BibitemShut {NoStop}%
\bibitem [{\citenamefont {Adamo}\ and\ \citenamefont
  {Barone}(1999)}]{adamo_toward_1999}%
  \BibitemOpen
  \bibfield  {author} {\bibinfo {author} {\bibfnamefont {C.}~\bibnamefont
  {Adamo}}\ and\ \bibinfo {author} {\bibfnamefont {V.}~\bibnamefont {Barone}},\
  }\href {\doibase 10.1063/1.478522} {\bibfield  {journal} {\bibinfo  {journal}
  {J. Chem. Phys.}\ }\textbf {\bibinfo {volume} {110}},\ \bibinfo {pages}
  {6158} (\bibinfo {year} {1999})}\BibitemShut {NoStop}%
\bibitem [{\citenamefont {Weigend}\ and\ \citenamefont
  {Ahlrichs}(2005)}]{weigend_balanced_2005}%
  \BibitemOpen
  \bibfield  {author} {\bibinfo {author} {\bibfnamefont {F.}~\bibnamefont
  {Weigend}}\ and\ \bibinfo {author} {\bibfnamefont {R.}~\bibnamefont
  {Ahlrichs}},\ }\href {\doibase 10.1039/b508541a} {\bibfield  {journal}
  {\bibinfo  {journal} {Phys. Chem. Chem. Phys.}\ }\textbf {\bibinfo {volume}
  {7}},\ \bibinfo {pages} {3297} (\bibinfo {year} {2005})}\BibitemShut
  {NoStop}%
\bibitem [{\citenamefont {Ferrer}\ and\ \citenamefont
  {Santoro}(2012)}]{ferrer_comparison_2012}%
  \BibitemOpen
  \bibfield  {author} {\bibinfo {author} {\bibfnamefont {F.~J.~A.}\
  \bibnamefont {Ferrer}}\ and\ \bibinfo {author} {\bibfnamefont
  {F.}~\bibnamefont {Santoro}},\ }\href {\doibase 10.1039/C2CP41169E}
  {\bibfield  {journal} {\bibinfo  {journal} {Phys. Chem. Chem. Phys.}\
  }\textbf {\bibinfo {volume} {14}},\ \bibinfo {pages} {13549} (\bibinfo {year}
  {2012})}\BibitemShut {NoStop}%
\bibitem [{\citenamefont {Worth}\ \emph {et~al.}()\citenamefont {Worth},
  \citenamefont {Beck}, \citenamefont {J{\"a}ckle}, \citenamefont {Vendrell},\
  and\ \citenamefont {Meyer}}]{MLMctdhpackageBla}%
  \BibitemOpen
  \bibfield  {author} {\bibinfo {author} {\bibfnamefont {G.~A.}\ \bibnamefont
  {Worth}}, \bibinfo {author} {\bibfnamefont {M.~H.}\ \bibnamefont {Beck}},
  \bibinfo {author} {\bibfnamefont {A.}~\bibnamefont {J{\"a}ckle}}, \bibinfo
  {author} {\bibfnamefont {O.}~\bibnamefont {Vendrell}}, \ and\ \bibinfo
  {author} {\bibfnamefont {H.-D.}\ \bibnamefont {Meyer}},\ }\href@noop {}
  {}\bibinfo {howpublished} {The MCTDH Package, Version 8.5.4 {S}ee
  http://mctdh.uni-hd.de/}\BibitemShut {NoStop}%
\bibitem [{\citenamefont {Press}\ \emph {et~al.}(2007)\citenamefont {Press},
  \citenamefont {Teukolsky}, \citenamefont {Vetterling},\ and\ \citenamefont
  {Flannery}}]{NumericalRecipes}%
  \BibitemOpen
  \bibfield  {author} {\bibinfo {author} {\bibfnamefont {W.~H.}\ \bibnamefont
  {Press}}, \bibinfo {author} {\bibfnamefont {S.~A.}\ \bibnamefont
  {Teukolsky}}, \bibinfo {author} {\bibfnamefont {W.~T.}\ \bibnamefont
  {Vetterling}}, \ and\ \bibinfo {author} {\bibfnamefont {B.~P.}\ \bibnamefont
  {Flannery}},\ }\href@noop {} {\emph {\bibinfo {title} {Numerical Recipes 3rd
  Edition: The Art of Scientific Computing}}},\ \bibinfo {edition} {3rd}\ ed.\
  (\bibinfo  {publisher} {Cambridge University Press},\ \bibinfo {address} {New
  York, NY, USA},\ \bibinfo {year} {2007})\BibitemShut {NoStop}%
\bibitem [{\citenamefont {Huh}\ and\ \citenamefont
  {Berger}(2011)}]{huh_application_2011}%
  \BibitemOpen
  \bibfield  {author} {\bibinfo {author} {\bibfnamefont {J.}~\bibnamefont
  {Huh}}\ and\ \bibinfo {author} {\bibfnamefont {R.}~\bibnamefont {Berger}},\
  }\href@noop {} {\bibfield  {journal} {\bibinfo  {journal} {Faraday Discuss.}\
  }\textbf {\bibinfo {volume} {150}},\ \bibinfo {pages} {363} (\bibinfo {year}
  {2011})}\BibitemShut {NoStop}%
\bibitem [{\citenamefont {Köppel}, \citenamefont {Domcke},\ and\ \citenamefont
  {Cederbaum}(2004)}]{cederbaum_lvc_2004}%
  \BibitemOpen
  \bibfield  {author} {\bibinfo {author} {\bibfnamefont {H.}~\bibnamefont
  {Köppel}}, \bibinfo {author} {\bibfnamefont {W.}~\bibnamefont {Domcke}}, \
  and\ \bibinfo {author} {\bibfnamefont {L.~S.}\ \bibnamefont {Cederbaum}},\
  }in\ \href@noop {} {\emph {\bibinfo {booktitle} {Conical Intersections}}},\
  \bibinfo {editor} {edited by\ \bibinfo {editor} {\bibfnamefont
  {W.}~\bibnamefont {Domcke}}, \bibinfo {editor} {\bibfnamefont {D.~R.}\
  \bibnamefont {Yarkony}}, \ and\ \bibinfo {editor} {\bibfnamefont
  {H.}~\bibnamefont {Köppel}}}\ (\bibinfo  {publisher} {World Scientific
  Co.},\ \bibinfo {year} {2004})\ pp.\ \bibinfo {pages} {323--368}\BibitemShut
  {NoStop}%
\end{thebibliography}
\end{document}